\documentclass[preprintnumbers,nofootinbib,noshowpacs,eqsecnum,prd,superscriptaddress,letterpaper]{revtex4-2}
\pdfoutput=1
\usepackage{graphicx}
\usepackage{amsmath,amssymb}
\usepackage{url}
\usepackage{multirow}
\usepackage{hyperref,color}
\usepackage{slashed}
\usepackage{rotating}
\usepackage{afterpage}
\usepackage{ifthen}
\usepackage{mciteplus}
\usepackage{multirow}
\usepackage{color}
\usepackage{pifont}
\usepackage{float}
\newcommand{\Frac}[2]{\frac{\displaystyle #1}{\displaystyle #2}}

\def\lsim{\raise0.3ex\hbox{$\;<$\kern-0.75em\raise-1.1ex\hbox{$\sim\;$}}}
\def\gsim{\raise0.3ex\hbox{$\;>$\kern-0.75em\raise-1.1ex\hbox{$\sim\;$}}}

\renewcommand{\baselinestretch}{1.2}

\newcommand{\agl}[2]{\langle #1 #2 \rangle}
\newcommand{\sqr}[2]{\lbrack #1 #2 \rbrack}
\newcommand{\hf}[0]{\tfrac{1}{2}}

\begin{document}
\renewcommand{\baselinestretch}{1.5}

\preprint{YITP-SB-2026-01}
\title{Unitarity limits on  triple gauge boson production}

\author{O.\ J.\ P.\ \'Eboli}
\email{eboli@if.usp.br}
\affiliation{Instituto de F\'{\i}sica, 
Universidade de S\~ao Paulo, S\~ao Paulo -- SP  05508-090, Brazil.}

\author{M.~C.~Gonzalez-Garcia}
\email{concha.gonzalez-garcia@stonybrook.edu}
\affiliation{C.N. Yang Institute for Theoretical Physics,
  Stony Brook University, Stony Brook New York 11794-3849, USA}
\affiliation{Departament de Fis\'{\i}ca Qu\`antica i
  Astrof\'{\i}sica and Institut de Ciencies del Cosmos, Universitat de
  Barcelona, Diagonal 647, E-08028 Barcelona, Spain}
\affiliation{Instituci\'o Catalana de Recerca i Estudis
  Avan\c{c}ats (ICREA), Pg. Lluis Companys 23, 08010 Barcelona,
  Spain.}
%

\begin{abstract}

  We derive the unitarity constraints on anomalous quartic gauge
  couplings originating from the $f \bar{f}^\prime \to V V' V''$
  channel with $V^{(')('')}=\gamma$, $Z$, $W^\pm$, and $H$. We also
  assess the importance of these bounds on the present and future
  experimental searches.

\end{abstract}


\maketitle

\section{Introduction}

The production of electroweak gauge bosons provides a window into
beyond the Standard Model (SM) physics at the CERN Large Hadron
Collider (LHC) because triple (TGC) and quartic (QGC) gauge-boson
couplings are determined by the SM gauge symmetry. Presently, the
studies of QGC are carried out in the exclusive production of
two~\cite{ATLAS:2016lse, CMS:2022dmc} or three~\cite{ATLAS:2017bon,
  ATLAS:2022xnu, ATLAS:2022wmu, ATLAS:2024nab, ATLAS:2025wyo,
  CMS:2014cdf, CMS:2021jji, CMS:2023rcv, CMS:2025oey} gauge bosons as
well as in the vector-boson-scattering production of electroweak gauge
boson pairs~\cite{ATLAS:2016snd, ATLAS:2019thr, ATLAS:2023dkz,
  CMS:2025dbm}. \smallskip

Up to this moment, there is no indication of the existence of any new
state or departure from the SM dynamics. Hence, it is natural to
assume that there is a mass gap between the ultraviolet extension of
the SM and the electroweak scale.  In this scenario, new physics
effects can be parametrized using effective field theory (EFT) and
they manifest themselves in the tails of kinematic distributions.  In
general, the most stringent constraints on Wilson coefficients of
effective operators that contain both TGC and QGC originate from the
study of their TGC component in the pair production of gauge
bosons. Therefore, most of the present LHC searches for effects of QGC
focus on the so-called genuine QGC operators -- that is, operators
that generate QGC but do not have any TGC associated with them.
\smallskip

It is well known that scattering amplitudes containing EFT operators
grow as the center-of-mass energy increases, eventually, presenting
violation of partial-wave unitarity. This indicates that the
perturbative use of EFT has a limited validity range.  Thus, when
probing anomalous QGC one must verify whether perturbative
partial-wave unitarity is satisfied in the signal simulations to
guarantee consistency of the analyses. Previously, unitarity
constraints on genuine QGC operators were obtained analyzing
two-to-two scattering of electroweak gauge bosons, here on denoted as
$V V \to V V$~\cite{Almeida:2020ylr, Arnold:2008rz}.  In this work we
complement the existing literature on the subject by systematically
deriving the unitarity limits on QGC originating from the
$f \bar{f}^\prime \to V V' V''$ channel with $V^{(')('')}=\gamma$,
$Z$, $W^\pm$, and $H$, here on denoted as $FF\to VVV$.  \smallskip

This paper is organized as follows: in Section~\ref{sec:lag} we
present the QCG operators that we consider in our analyses, as well as
the basic expressions of required for the three-particle partial-wave
projections needed for our studies. Section~\ref{sec:results} contains
our results using the unitarity violating partial-wave projections
presented in Appendix~\ref{app:hel} while Section~\ref{sec:disc} is
dedicated to the discussion of the impact of the $FF \to VVV$
unitarity limits on present and future experimental analyses.

\section{Theoretical framework}
\label{sec:lag}

\subsection{Effective field theory}

Low-energy EFTs are based on the particle content as well as the
symmetries of theory. Here, we assume that the observed Higgs boson is
the SM one, {\em i.e.} it belongs to an electroweak scalar doublet.
Therefore, we can construct a low-energy effective theory where the
$SU(2)_L \otimes U(1)_Y$ gauge symmetry is linearly
realized~\cite{Buchmuller:1985jz, Leung:1984ni, DeRujula:1991ufe,
  Hagiwara:1993ck, GonzalezGarcia:1999fq, Grzadkowski:2010es,
  Passarino:2012cb} which takes the form
\begin{equation}
{\cal L}_{\rm eff} = {\cal L}_{\text SM} + \sum_{n=5}^\infty\sum_i
\frac{f^{(n)}_i}{\Lambda^{n-4}} {\cal O}^{(n)}_i \;\; ,
\label{l:eff}
\end{equation}
where the dimension--n operators ${\cal O}^{(n)}_i$ contains the SM
fields and their covariant derivatives. We denote by $f^{(n)}_i$ the
corresponding Wilson coefficient and by $\Lambda$ the ultraviolet
energy scale.  The lowest dimension containing genuine QGC operators
is eight~\cite{Eboli:2006wa, Eboli:2016kko, Durieux:2024zrg}. Assuming
$C$ and $P$ conservation, there are 20 dimension-eight genuine QGC
operators that we classify by their number of gauge-boson field
strengths. There are 3 QGC operators formed of four covariant
derivatives of the Higgs field
\begin{equation}
\begin{array}{lll}
  {\cal O}_{S,0} = 
\left [ \left ( D_\mu \Phi \right)^\dagger
 D_\nu \Phi \right ] \times 
\left [ \left ( D^\mu \Phi \right)^\dagger
D^\nu \Phi \right ]
&\hbox{   ,   }
&
{\cal O}_{S,1} =
 \left [ \left ( D_\mu \Phi \right)^\dagger
 D^\mu \Phi  \right ] \times
\left [ \left ( D_\nu \Phi \right)^\dagger
D^\nu \Phi \right ]
\; , 
\\
&&
\\
  {\cal O}_{S,2} =
 \left [ \left ( D_\mu \Phi \right)^\dagger
 D_\nu \Phi  \right ] \times
\left [ \left ( D^\nu \Phi \right)^\dagger
D^\mu \Phi \right ]
\; ,
\end{array}
\label{eq:dphi}
\end{equation}
where $\Phi$ stands for the Higgs doublet, the covariant derivative is
given by
$D_\mu \Phi = (\partial_\mu + i g W^j_\mu \frac{\sigma^j}{2} + i
g^\prime B_\mu \frac{1}{2}) \Phi$ and $\sigma^j$ ($j=1,2,3$) represent
the Pauli matrices. In addition, there are 7 QGC operators that
exhibit two covariant derivatives of the Higgs field as well as two
field strengths:
\begin{equation}
\begin{array}{lcll}
 {\cal O}_{M,0} =   \hbox{Tr}\left [ \widehat{W}_{\mu\nu} \widehat{W}^{\mu\nu} \right ]
\times  \left [ \left ( D_\beta \Phi \right)^\dagger
D^\beta \Phi \right ]
&,& 
 {\cal O}_{M,1} 
=   \hbox{Tr}\left [ \widehat{W}_{\mu\nu} \widehat{W}^{\nu\beta} \right ]
\times  \left [ \left ( D_\beta \Phi \right)^\dagger
D^\mu \Phi \right ]
&,
\\
 {\cal O}_{M,2} =   \left [ B_{\mu\nu} B^{\mu\nu} \right ]
\times  \left [ \left ( D_\beta \Phi \right)^\dagger
D^\beta \Phi \right ]
&,&
 {\cal O}_{M,3} =   \left [ B_{\mu\nu} B^{\nu\beta} \right ]
\times  \left [ \left ( D_\beta \Phi \right)^\dagger
D^\mu \Phi \right ]
&,
\\
  {\cal O}_{M,4} = \left [ \left ( D_\mu \Phi \right)^\dagger \widehat{W}_{\beta\nu}
 D^\mu \Phi  \right ] \times B^{\beta\nu}
&,&
  {\cal O}_{M,5} = \left [ \left ( D_\mu \Phi \right)^\dagger \widehat{W}_{\beta\nu}
 D^\nu \Phi  \right ] \times B^{\beta\mu}+ {\rm h.c.}
&,
\\
  {\cal O}_{M,7} = \left [ \left ( D_\mu \Phi \right)^\dagger \widehat{W}_{\beta\nu}
\widehat{W}^{\beta\mu} D^\nu \Phi  \right ]  
&,&&
\end{array}
\label{eq:lind2}
\end{equation}
where
$\widehat{W}_{\mu\nu} \equiv W^j_{\mu\nu} \frac{\sigma^j}{2}$ is the
$SU(2)_L$ field strength while $B_{\mu\nu}$ stands for the $U(1)_Y$
one. Finally, we can construct 10 operators with 4 field strengths:
\begin{equation}
\begin{array} {lcl}
 {\cal O}_{T,0} =   \hbox{Tr}\left [ \widehat{W}_{\mu\nu} \widehat{W}^{\mu\nu} \right ]
\times   \hbox{Tr}\left [ \widehat{W}_{\alpha\beta} \widehat{W}^{\alpha\beta} \right ]
&, &
 {\cal O}_{T,1} =   \hbox{Tr}\left [ \widehat{W}_{\alpha\nu} \widehat{W}^{\mu\beta} \right ]
\times   \hbox{Tr}\left [ \widehat{W}_{\mu\beta} \widehat{W}^{\alpha\nu} \right ]
\\
 {\cal O}_{T,2} =   \hbox{Tr}\left [ \widehat{W}_{\alpha\mu} \widehat{W}^{\mu\beta} \right ]
\times   \hbox{Tr}\left [ \widehat{W}_{\beta\nu} \widehat{W}^{\nu\alpha} \right ]
&,&
         {\cal O}_{T,3}
         =  \hbox{Tr}\left [ \widehat{W}_{\mu\nu} \widehat{W}_{\alpha\beta} \right ]
\times   \hbox{Tr}\left [ \widehat{W}^{\alpha\nu} \widehat{W}^{\mu\beta} \right ]
 \\
{\cal O}_{T,4} =   \hbox{Tr}\left [ \widehat{W}_{\mu\nu} \widehat{W}_{\alpha\beta} \right ]
\times   B^{\alpha\nu} B^{\mu\beta}
 &,&
 {\cal O}_{T,5} =   \hbox{Tr}\left [ \widehat{W}_{\mu\nu} \widehat{W}^{\mu\nu} \right ]
\times   B_{\alpha\beta} B^{\alpha\beta}
\\ 
 {\cal O}_{T,6} =   \hbox{Tr}\left [ \widehat{W}_{\alpha\nu} \widehat{W}^{\mu\beta} \right ]
\times   B_{\mu\beta} B^{\alpha\nu} 
&,&
 {\cal O}_{T,7} =   \hbox{Tr}\left [ \widehat{W}_{\alpha\mu} \widehat{W}^{\mu\beta} \right ]
\times   B_{\beta\nu} B^{\nu\alpha} 
\\ 
 {\cal O}_{T,8} =   B_{\mu\nu} B^{\mu\nu}  B_{\alpha\beta} B^{\alpha\beta}
&,& 
 {\cal O}_{T,9} =  B_{\alpha\mu} B^{\mu\beta}   B_{\beta\nu} B^{\nu\alpha} 
\; . 
\end{array}
\label{eq:lan-t}
\end{equation}

The above  20 genuine QGC operators give rise to modifications to vertices
$VVVV$, $VVVH$ and $VVHH$ ($V=W^\pm$, $Z$ and $A$) that are compatible
with electric charge, $C$ and $P$ conservation~\cite{Eboli:2016kko}.\smallskip

\subsection{Three-particle partial waves}

In order to obtain the unitarity bounds from the process leading to
the production of three electroweak gauge bosons
\begin{equation}
f_{1,\lambda_1}\bar f_{2,\lambda_2}\rightarrow
V_{3,\lambda_3}V_{4,\lambda_4}V_{5,\lambda_5}
\label{eq:ffvvv}
\end{equation}
we project the corresponding helicity amplitude into a kinematic basis
$\left\{ {\cal} B^{J,a}_{\lambda_1,\lambda_2; \lambda_3,
    \lambda_4,\lambda_5} \right\}$ with definite total angular
momentum $J$. The index $a$ represents additional quantum numbers
needed to completely characterize the states besides helicities and
total angular momentum. The partial wave amplitudes
$T^{J,a} (f_{1,\lambda_1} \bar f_{2,\lambda_2}\rightarrow
V_{3,\lambda_3} V_{4,\lambda_4}V_{5,\lambda_5})$ are obtained by
expanding the scattering amplitude is this basis:
\begin{equation}
  {\cal M}
    (f_{1,\lambda_1}\bar f_{2,\lambda_2}\rightarrow
    V_{3,\lambda_3}V_{4,\lambda_4}V_{5,\lambda_5})
    =\sum_{J,a}
    T^{J,a} (f_{1,\lambda_1}\bar f_{2,\lambda_2}\rightarrow
    V_{3,\lambda_3}V_{4,\lambda_4}V_{5,\lambda_5}) \times 
     {\cal B}^{J,a}_{\lambda_1,\lambda_2; \lambda_3
       \lambda_4\lambda_5}  \;.
\end{equation}

It is convenient to normalize the partial-wave basis such that
\begin{equation}
  \int d\Phi_{f_1,f_2} d\Phi_{V_3,V_4,V_5}~
  {\cal B}^{J,a}_{\lambda_1\lambda_2; \lambda_3, \lambda_4,\lambda_5}
    ({\cal B}^{J',b}_{\lambda_1,\lambda_2; \lambda_3, \lambda_4,\lambda_5})^*=
      (2J+1) \delta^{J,J'} \delta ^{a,b} \;,
\end{equation}
where $\Phi_S$ is the phase space for states $S$. The normalization
allows us to write the partial-wave amplitudes as
\begin{equation}
 T^{J,a} (f_{1,\lambda_1}\bar f_{2,\lambda_2}\rightarrow
 V_{3,\lambda_3}V_{4,\lambda_4}V_{5,\lambda_5})
 = \frac{1}{2J+1}\int d\Phi_{f_1,f_2} d\Phi_{V_3,V_4,V_5}
  {\cal M}_{\lambda_1\lambda_2; \lambda_3, \lambda_4,\lambda_5}
  ({\cal B}^{J,a}_{\lambda_1,\lambda_2; \lambda_3,
    \lambda_4,\lambda_5})^* \;.
\end{equation}

The partial-wave basis can be constructed from the kinematic
polynomials obtained using spinor-helicity
techniques~\cite{Bresciani:2025toe}.  The kinematic polynomials
${\cal P}^{J,a}_{\lambda_1,\lambda_2; \lambda_3 \lambda_4\lambda_5}$
relevant for $2\rightarrow 3$ processes are given in the Appendix A of
Ref.~\cite{Bresciani:2025toe}.  For the process~\eqref{eq:ffvvv} with
massless fermions the relevant polynomials are
\begin{equation}
  \begin{array}{cccl} 
  \lambda_1,\lambda_2;\lambda_3,\lambda_4,\lambda_5 &\,\,\,\,\, J\,\,\,\,\,
  & \,\,\,\,\, a\,\,\,\,\, & {\cal P}^{J,a}\\
  \hline\hline
  (-\hf,\hf;0,0,0) & 1 & 1 &  192\sqrt{2}\pi^2S^{-3/2} \agl{1}{4}\sqr{4}{2} \\
    &1&2& 192\sqrt{2}\pi^2S^{-3/2} \agl{1}{5}\sqr{5}{2} \\\hline
  (-\hf,\hf;0,0,1) & 1 & 1 &
  384\sqrt{\tfrac{5}{7}}\pi^2S^{-2} \agl{1}{3}\sqr{5}{2}\sqr{5}{3} \\
  &1 &2 & 384\sqrt{\tfrac{5}{7}}\pi^2S^{-2} \agl{1}{4}\sqr{5}{2}\sqr{5}{4}
  \\\hline
  (-\hf,\hf;0,1,1) & 1 & 1 & 576\pi^2S^{-5/2} \agl{1}{5}
  \sqr{5}{2}\sqr{5}{4}^2  \\
    &1 & 2& 576\pi^2S^{-5/2} \agl{1}{4}\sqr{4}{2}\sqr{5}{4}^2 \\
  &1 & 3 & 288\sqrt{10}\pi^2S^{-5/2} \agl{1}{3}\sqr{4}{2}
  \sqr{5}{3}\sqr{5}{4} \\\hline
  (-\hf,\hf;-1,0,1)
  & 1 & 1 &288\sqrt{10}\pi^2S^{-5/2}
  \agl{1}{3}\agl{3}{4}\sqr{5}{2}\sqr{5}{4} \\ \hline
  (-\hf,\hf;-1,1,1)
  & 1 &  1 & 1152\sqrt{\tfrac{35}{31}}
  \pi^2S^{-3} \agl{1}{3}\agl{3}{5}\sqr{5}{2}\sqr{5}{4}^2 \\
  &1&2& 1152\sqrt{\tfrac{35}{31}}\pi^2S^{-3}
  \agl{1}{3}\agl{3}{4}\sqr{4}{2}\sqr{5}{4}^2 \\ \hline
  (-\hf,\hf;1,1,1) & 1 & 1 & 1152\sqrt{\tfrac{35}{31}}\pi^2S^{-3}
  \agl{1}{4}\sqr{3}{2}\sqr{4}{3}\sqr{5}{4}^2 \\
  &1 & 2&  1152\sqrt{\tfrac{35}{31}}\pi^2S^{-3}
  \agl{1}{5}\sqr{3}{2}\sqr{5}{3}\sqr{5}{4}^2 \\
  &1 & 3 &  1152\sqrt{\tfrac{35}{31}}\pi^2S^{-3}
  \agl{1}{3}
  \sqr{4}{2}\sqr{4}{3}\sqr{5}{3}^2 \\
  &1 & 4 &1152\sqrt{\tfrac{35}{31}}\pi^2S^{-3}
  \agl{1}{5}\sqr{4}{2}\sqr{5}{3}^2\sqr{5}{4} \\
  &1 & 5& 576\sqrt{14}\pi^2S^{-3} \agl{1}{4}
  \sqr{4}{2}\sqr{4}{3}\sqr{5}{3}\sqr{5}{4}
  \\\hline\hline
  \end{array}
\label{eq:basis}  
\end{equation}
where $S=(p_1+p_2)^2$. In addition, the polynomials with
$\lambda_i\rightarrow -\lambda_i$ are obtained with the substitution
$\agl{i}{j}\leftrightarrow \sqr{j}{i}$ for all $i,j$. We use the
conventions for the square and angle brackets of
Ref.~\cite{Bresciani:2025toe} while the phase space integrals where
carried out using spinor helicity variables as defined in the Appendix
A of Ref.~\cite{EliasMiro:2020tdv}. \smallskip

The kinematic polynomials in Eq.~\eqref{eq:basis} are normalized to
one, however, they are not orthogonal for different values of the
additional quantum numbers ``$a$''.  Hence, to define the partial-wave
amplitudes we construct the corresponding normalized and orthogonal
kinematic basis as
\begin{equation}
  {\cal B}^{J,a}_{\lambda_1\lambda_2; \lambda_3, \lambda_4,\lambda_5}
  =\sum_{b} {\cal R}^{J}_{ab} \times
  {\cal P}^{J,b}_{\lambda_1\lambda_2; \lambda_3, \lambda_4,\lambda_5}
  \label{eq:rotation}
\end{equation}
where the ``rotation'' matrices are given by
\begin{eqnarray}
  {\cal R}^{1}_{-\hf,\hf;000}=\left(\begin{array}{cc}
    1 & 1 \\
      \tfrac{1}{\sqrt{3}} &-\tfrac{1}{\sqrt{3}}
    \end{array}\right)\;,
\;\;\;
    {\cal R}^{1}_{-\hf,\hf;001}=\frac{1}{2}\left(\begin{array}{cc}
      \sqrt{\tfrac{7}{5}} &\sqrt{\tfrac{7}{5}} \\
      \sqrt{\tfrac{7}{2}} &-\sqrt{\tfrac{7}{2}}
    \end{array}\right)
   \;, \;\;\;
        {\cal R}^{1}_{-\hf,\hf;011}=\left(\begin{array}{ccc}
          0.442&0.405&0.317\\0.175 &-0.596 & 1.00 \nonumber\\
          2.25 & 1.89 & 0.730
                                          \end{array}\right) \\
  \\ \nonumber
{\cal R}^{1}_{-\hf,\hf;-1,1,1}=\tfrac{1}{2}\left(
    \begin{array}{cc}
      \sqrt{\tfrac{31}{21}} &\sqrt{\tfrac{31}{21}}\\
      \sqrt{\tfrac{31}{10}} & -\sqrt{\tfrac{31}{10}}
    \end{array}\right)
   \;\;,\;\;\; {\cal R}^{1}_{-\hf,\hf;1,1,1}=\left(\begin{array}{ccccc}
 0.203 & 0.429& -0.0671& 0.372& 
  0.280\\0.320& -0.0197& 0.606& -0.331& 
  0.384 \\-0.940& 0.125& 0.431& 0.220& 
  0.300 \\-0.174& -0.361& -0.781& -0.396& 
  1.019 \\0.684& -2.84& 0.597& 2.58& 0.572\end{array}\right) \;.
\end{eqnarray}  
We can then construct the partial-wave amplitudes as the corresponding
linear combination
\begin{equation}
T^{J,a} (f_{1,\lambda_1}\bar f_{2,\lambda_2}\rightarrow
 V_{3,\lambda_3}V_{4,\lambda_4}V_{5,\lambda_5})
 = \sum_b ({\cal R}^{J}_{ab})^* \times
 \tilde{T}^{J,b} (f_{1,\lambda_1}\bar f_{2,\lambda_2}\rightarrow
 V_{3,\lambda_3}V_{4,\lambda_4}V_{5,\lambda_5})
\end{equation}
in terms of the projections over the non-orthogonal basis
\begin{equation}
 \tilde{T}^{J,b} (f_{1,\lambda_1}\bar f_{2,\lambda_2}\rightarrow
 V_{3,\lambda_3}V_{4,\lambda_4}V_{5,\lambda_5})=  
 \frac{1}{2J+1}\int d\Phi_{f_1,f_2}
 d\Phi_{V_3,V_4,V_5}
  {\cal M}_{\lambda_1\lambda_2; \lambda_3, \lambda_4,\lambda_5}
  ({\cal P}^{J,b}_{\lambda_1,\lambda_2; \lambda_3, \lambda_4,\lambda_5})^*\;.
\end{equation}

Finally, the optical theorem can be generalized leading to
\begin{equation}
  T^{J,a}_{i\to f} - \left( T^{J,a}_{i\to f}\right)^\star = i \sum_X T^{J,a}_{i\to
    X} \left(    T^{J,a}_{f\to     X} \right)^\star 
\end{equation}
that leads to the partial-wave unitarity bound
\begin{equation}
|T^{J,a} (f_{1,\lambda_1}\bar f_{2,\lambda_2}\rightarrow
    V_{3,\lambda_3}V_{4,\lambda_4}V_{5,\lambda_5})|<1 \;.
\end{equation}

\section{Unitarity limits on genuine QGC}
\label{sec:results}

\subsection{Summary of unitarity limits from $VV \to VV$}

Ref.~\cite{Almeida:2020ylr} contains a detailed study of unitarity
limits on QGCs stemming from $VV \to VV$ scattering processes. These
bounds were obtained from the $J=0$ and $J=1$ partial-wave amplitudes
for the scattering of gauge and Higgs bosons.  Considering just one
operator at a time the tightest bounds arise from the analysis of the
$J=0$ partial wave whose fastest energy growing behaviour is of the
form
\begin{equation}
  T^{J=0}={\rm C}^{VV}_{i}
  \times \frac{f_i}{\Lambda^4} \, S_{VV}^2\;,
\end{equation}
where ${\rm C}^{VV}_{X}$ are numerical constants and $\sqrt{S_{VV}}$
stands for the diboson center-of-mass energy.  We found the largest
coefficients for one operator at a time to be
\begin{eqnarray}
&&  |{\rm C}^{VV}_{S,i,\rm{max}}|=\left(\frac{1}{32 \pi}, \frac{7}{96 \pi},\frac{5}{96 \pi}
  \right)\, \hspace*{6.5cm} \mbox{for $i=0,1,2$}
  \nonumber\\
 && |{\rm C}^{VV}_{M,i,{\rm max}}|=
        \left(\frac{\sqrt{6}}{32 \pi},
                            \frac{\sqrt{6}}{128 \pi},
                            \frac{\sqrt{2}}{16 \pi},
                            \frac{\sqrt{2}}{64 \pi},
                            \frac{1}{32 \pi},
                            \frac{1}{64 \pi},
                            \frac{\sqrt{6}}{256 \pi}\right)\;,
\hspace*{2.5cm} \mbox{for $i=0,1,2,3,4,5,7$}
  \label{eq:cvvmax}                          \\
&& |{\rm C}^{VV}_{T,i,{\rm max}}|=\left(\frac{5}{12 \pi},\frac{5}{24 \pi},\frac{13}{96 \pi},\frac{3}{32 \pi},\frac{1}{16 \pi},\frac{\sqrt{3}}{8 \pi},\frac{7}{48 \pi},\frac{\sqrt{3}}{32 \pi},\frac{2}{3 \pi},\frac{7}{24 \pi}
                            \right)
                            \hspace*{1cm} \mbox{for $i=0 \dots 9$}
                            \nonumber
\end{eqnarray}
We present the strongest
unitarity limits from these amplitudes in Table~\ref{tab:VV}. As a reference, for
$f_{X,i} /\Lambda^4=1$ TeV$^{-4}$ unitarity violation takes place at
diboson invariant masses, $M_{VV}$, in excess of 1.5-4.7 TeV depending
on the QGC operator.  \smallskip

\begin{table}
\begin{tabular}{|c|c||c|c|}
\hline
Coefficient& Bound & Coefficient & Bound\\\hline
$\left | \Frac{f_{S,0}}{\Lambda^4} \right|$& $32\pi \Frac{\rm TeV^4}{S_{VV}^2}= 100 \Frac{\rm TeV^4}{S_{VV}^2}$&
$\left | \Frac{f_{T,0}}{\Lambda^4} \right|$& $\Frac{12\pi}{5}\Frac{\rm TeV^4}{S_{VV}^2}=7.5   \Frac{\rm TeV^4}{S_{VV}^2}$\\
$\left | \Frac{f_{S,1}}{\Lambda^4} \right|$& $\Frac{96\pi}{7}
                                             \Frac{\rm
                                             TeV^4}{S_{VV}^2}={ 43} \Frac{\rm TeV^4}{S_{VV}^2}$&
$\left | \Frac{f_{T,1}}{\Lambda^4} \right|$&$ \Frac{24\pi}{5} \Frac{\rm TeV^4}{S_{VV}^2}=15   \Frac{\rm TeV^4}{S_{VV}^2}$\\
$\left | \Frac{f_{S,2}}{\Lambda^4} \right|$&$\Frac{96\pi}{5}\Frac{\rm TeV^4}{S_{VV}^2}= 60  \Frac{\rm TeV^4}{S_{VV}^2}$&
$\left | \Frac{f_{T,2}}{\Lambda^4} \right|$&$ \Frac{96\pi}{13} \Frac{\rm TeV^4}{S_{VV}^2}= 23  \Frac{\rm TeV^4}{S_{VV}^2}$\\
$\left | \Frac{f_{M,0}}{\Lambda^4} \right|$
&$ \Frac{32 \pi}{\sqrt{6}} \Frac{\rm TeV^4}{S_{VV}^2}=41   \Frac{\rm TeV^4}{S_{VV}^2}$&
$\left | \Frac{f_{T,3}}{\Lambda^4} \right|$&$ \Frac{32\pi}{3} \Frac{\rm TeV^4}{S_{VV}^2}= 33  \Frac{\rm TeV^4}{S_{VV}^2}$\\
$\left | \Frac{f_{M,1}}{\Lambda^4} \right|$&$  \Frac{128 \pi}{\sqrt{6}}\Frac{\rm TeV^4}{S_{VV}^2}= 164  \Frac{\rm TeV^4}{S_{VV}^2}$&
$\left | \Frac{f_{T,4}}{\Lambda^4} \right|$&$ 16\pi \Frac{\rm TeV^4}{S_{VV}^2}=50   \Frac{\rm TeV^4}{S_{VV}^2}$\\
$\left | \Frac{f_{M,2}}{\Lambda^4} \right|$&$  \Frac{16 \pi}{\sqrt{2}}\Frac{\rm TeV^4}{S_{VV}^2}= 35  \Frac{\rm TeV^4}{S_{VV}^2}$&
$\left | \Frac{f_{T,5}}{\Lambda^4} \right|$&$ \Frac{8\pi}{\sqrt{3}} \Frac{\rm TeV^4}{S_{VV}^2}= 14  \Frac{\rm TeV^4}{S_{VV}^2}$\\
$\left | \Frac{f_{M,3}}{\Lambda^4} \right|$&$  \Frac{64 \pi}{\sqrt{2}}\Frac{\rm TeV^4}{S_{VV}^2}= 142  \Frac{\rm TeV^4}{S_{VV}^2}$&
$\left | \Frac{f_{T,6}}{\Lambda^4} \right|$&$  \Frac{48\pi}{7}\Frac{\rm TeV^4}{S_{VV}^2}=21   \Frac{\rm TeV^4}{S_{VV}^2}$\\
$\left | \Frac{f_{M,4}}{\Lambda^4} \right|$&$   32\pi \Frac{\rm TeV^4}{S_{VV}^2}= 100  \Frac{\rm TeV^4}{S_{VV}^2}$&
$\left | \Frac{f_{T,7}}{\Lambda^4} \right|$&$
                                             \Frac{32\pi}{\sqrt{3}}\Frac{\rm
                                             TeV^4}{S_{VV}^2}={ 58}  \Frac{\rm TeV^4}{S_{VV}^2}$\\
$\left | \Frac{f_{M,5}}{\Lambda^4} \right|$&$  64\pi \Frac{\rm TeV^4}{S_{VV}^2}= 201 \Frac{\rm TeV^4}{S_{VV}^2}$&
$\left | \Frac{f_{T,8}}{\Lambda^4} \right|$&$ \Frac{3\pi}{2} \Frac{\rm
                                             TeV^4}{S_{VV}^2}= { 4.7}  \Frac{\rm TeV^4}{S_{VV}^2}$\\
$\left | \Frac{f_{M,7}}{\Lambda^4} \right|$&$  \Frac{256 \pi}{\sqrt{6}}\Frac{\rm TeV^4}{S_{VV}^2}= 328  \Frac{\rm TeV^4}{S_{VV}^2}$&
$\left | \Frac{f_{T,9}}{\Lambda^4} \right|$&$  \Frac{24\pi}{7}\Frac{\rm TeV^4}{S_{VV}^2}= 11  \Frac{\rm TeV^4}{S_{VV}^2}$
\\\hline
\end{tabular}
\caption{Most stringent unitarity constraints from $VV\rightarrow VV$
  processes on the Wilson coefficients of the ${\cal O}_{X,j}$
  operators when just one coefficient is non vanishing.}
\label{tab:VV}
\end{table}  

\subsection{Unitarity bounds from $FF \rightarrow VVV$}

Many experimental studies of anomalous QGCs are carried out studying
the exclusive production of three electroweak gauge bosons, {\em i.e.}
$FF \rightarrow VVV$.  Therefore, we analyze the unitarity limits
associated to the production of three gauge and/or Higgs bosons. Here,
we focus on the leading processes that present the fastest growth as a
function of the three-gauge-boson center-of-mass energy
$\sqrt{S_{VVV}}$. \smallskip

\subsubsection{ ${\cal O}_{S,i}$ operators}

For operators ${\cal O}_{S,i}$, the partial-wave amplitudes displaying
the fastest increase with $\sqrt{S_{VVV}}$ are the ones with three
longitudinally polarized gauge bosons and they are shown in
Table~\ref{tab:000}. These results are presented for the
non-orthogonal partial-wave states~\footnote{For the sake of
  concreteness we show the results for initial left-handed 
  currents, that is, with {\sl incoming} helicities
  $\lambda_1=-\lambda_2=-\frac{1}{2}$ as they are assumed to be
  massless).
  Furthermore the results for initial right-handed fermion currents can
  be obtained by projecting over the parity flipped basis, and
  changing the chirality of the corresponding fermion weak charge
  involved in the process.}.  The high energy behaviour of the leading
helicity amplitudes is
\begin{equation}
  T^{1,a}_{-\frac{1}{2},\frac{1}{2};0,0,0}={\rm C}^{VVV}_{S,i}\times
  \frac{f_{S,i} }{\Lambda^4}\, S_{VVV}^{3/2}\, v \;,
\end{equation}
where $v$ stands for the Higgs vacuum expectation value.  This
high-energy behaviour is to be expected from simple dimensional analysis.
For ${\cal O}_{S,i}$ operators the four-vector-boson coupling originates
from the four Higgs covariant derivatives and therefore it is proportional
to $v^4$ while the presence of three polarized gauge bosons leads to a factor
$v^{-3}$ resulting in a scattering amplitude proportional to $v S$ 
since the scattering amplitude has inverse mass dimension.
Taking into account that the
  two-to-three phase space scales as $S_{VVV}$ and the partial-wave
  basis as $S_{VVV}^{-1/2}$, we obtain a partial-wave amplitude
  proportional to $v S_{VVV}^{3/2}$.   \smallskip

Explicitly, after rotating to the partial waves on the orthogonal
basis with Eq.~\eqref{eq:rotation}, we find the largest partial-wave
coefficients for each coupling to be
\begin{equation}
  \left|{\rm C}^{VVV}_{S,i,\rm max}\right|=\left(1.8\,,\,0.90\,,\,0.90 \right)\times 10^{-5}
\end{equation}
for $i=0,1,2$. They arise from the CC processes
$f_d\,\bar{f}_u \rightarrow W^- W^+ W^-$.  The corresponding unitarity
bounds read
\begin{eqnarray}
  \left| \Frac{f_{S,0}}{\Lambda^4} \right|\leq 2.3 \times 10^5\left(
  \Frac{\rm TeV^2}{S_{VVV}}\right)^{\frac{3}{2}}\; ,&
                                                \hspace*{0.5cm}
                                                \left|\Frac{f_{S,1}}{\Lambda^4} \right|\leq 4.5 \times 10^5\left(
                                                \Frac{\rm TeV^2}{S_{VVV}}\right)^{\frac{3}{2}}\; ,&
                                                                                              \hspace*{0.5cm}\left|\Frac{f_{S,2}}{\Lambda^4} \right|\leq 4.5\times
                                                                                              10^5\left(
                                                                                              \Frac{\rm TeV^2}{S_{VVV}}\right)^{\frac{3}{2}}\; .
\label{eq:boundsVVVS}  
\end{eqnarray}  
In summary, the $FF \to VVV$ amplitudes grow slower with $S_{VVV}$
than the corresponding ones from $VV\rightarrow VV$ on
$S_{VV}$. Furthermore they exhibit smaller numerical coefficients.
Consequently, the unitarity bounds on the $f_{S,i}$ Wilson
coefficients from triple gauge boson production,
Eq.~\eqref{eq:boundsVVVS}, are always weaker than those from
scattering of gauge boson pairs displayed in Table~\ref{tab:VV}.  For
instance, assuming that $f_{S,i}/\Lambda^4 = 1$ TeV$^{-4}$ unitarity
is violated in $FF\to VVV$ at the center-of-mass energies of (61, 76,
76) TeV for $i=0, 1, 2$, respectively, while it is violated in
$VV\to VV$ at the center-of-mass energies of (3.2, 2.8, 2.8) TeV for
$i=0, 1, 2$.  \smallskip

\subsubsection{ ${\cal     O}_{M,i}$ operators}

All ${\cal O}_{M,i}$ operators contain two field strengths, therefore,
their contributions to scattering amplitudes vanish when all vector
boson polarizations are longitudinal. In this case the amplitudes 
with fastest increase with $\sqrt{S_{VVV}}$ are the ones with two
longitudinally polarized and one transversely polarize gauge bosons. 
In this case the amplitude contains a
factor $v^2$ originating from the two covariant derivatives which cancels
out the factor $v^{-2}$ due to the two longitudinally polarized gauge bosons,
resulting in  a scattering amplitude that grows as $S_{VVV}^{3/2}$. Again,
taking into account that the
  two-to-three phase space scales as $S_{VVV}$ and the partial-wave
  basis as $S_{VVV}^{-1/2}$, we  
find   leading contributions of these operators depending  upon $S_{VVV}$ as
%
%
\begin{equation}
  T^{1,a}_{-\frac{1}{2},\frac{1}{2};0,0,1}={\rm C}^{VVV}_{M,i}\times
  \frac{f_{M,i}}{\Lambda^4} \,    S_{VVV}^2\;.
\end{equation}
%
%
The ${\cal O}_{M,i}$ non-orthogonal partial-wave amplitudes for the
relevant processes are presented in Table ~\ref{tab:VVVfm}. Using
these results and Eq.~\eqref{eq:rotation}, we can evaluate the fastest
growing partial wave for each Wilson coefficient:
\begin{equation}
  \left|
      {\rm C}^{VVV}_{M,i,\rm max}\right|  
  =\left( 2.3\,,\,1.6\,,\,2.3 \,,\, 1.5 \,,\,1.5 \,,\,0.97  \,,\,0.78
  \right)\times 10^{-5}\;
 \label{eq:cvvvM}   
\end{equation}
for $i=0,1,2,3,4,5,7$, respectively.  They arise from the CC processes
$f_d\,\bar{f}_u \rightarrow W^- W^+ W^-,Z Z W^-, HH W^-$ for coupling
$i=0,1,7$, from CC processes
$f_d\,\bar{f}_u \rightarrow Z W^- \gamma, H W^- \gamma$ for coupling
$i=4,5$, and from NC processes $\ell^- \ell^+\rightarrow W^+W^-\gamma$
and $\nu_\ell \bar\nu_\ell\rightarrow W^+W^-\gamma$ for $i=2,3$.  The
corresponding unitarity bounds are
\begin{eqnarray}
&&  \left | \Frac{f_{M,0}}{\Lambda^4} \right|< 4.2 \times { 10^{4}} \Frac{\rm TeV^4}{S_{VVV}^2}\;,
  \hspace*{2cm} 
  \left | \Frac{f_{M,1}}{\Lambda^4} \right|< 6.4 \times { 10^{4}} \Frac{\rm TeV^4}{S_{VVV}^2}\;,\nonumber\\
&&  \left | \Frac{f_{M,2}}{\Lambda^4} \right|< 4.4 \times { 10^{4}} \Frac{\rm TeV^4}{S_{VVV}^2}\;,
  \hspace*{2cm} 
  \left | \Frac{f_{M,3}}{\Lambda^4} \right|< 6.6 \times { 10^{4}} \Frac{\rm TeV^4}{S_{VVV}^2}\;,\nonumber\\
&&  \left | \Frac{f_{M,4}}{\Lambda^4} \right|< {6.8} \times { 10^{4}} \Frac{\rm TeV^4}{S_{VVV}^2}\;,
  \hspace*{2cm} 
  \left | \Frac{f_{M,5}}{\Lambda^4} \right|< 1.0 \times { 10^{5}} \Frac{\rm TeV^4}{S_{VVV}^2}\;,\\
&&  \left | \Frac{f_{M,7}}{\Lambda^4} \right|< 1.3 \times { 10^{5}} \Frac{\rm TeV^4}{S_{VVV}^2}\;.
\nonumber
\end{eqnarray}  
We can see from the last limits that despite the growth of the
scattering amplitudes presenting the same functional dependence on the
center-of-mass energy as those for $VV \to VV$, the corresponding
numerical coefficients are smaller by 2-3 orders of magnitude.  This
is so because of the higher partial wave ($J=1$) and larger
normalization factors of partial waves for three-body final states as
well as from the additional gauge coupling constant in the $ffV$
vertex. This renders the unitarity constraints on the Wilson
coefficients $f_{M,i}$ from the triple gauge boson production
substantially weaker than those from gauge-boson two-to-two
scattering.  We can assess that impact of such small numerical
coefficients by obtaining the center-of-mass energy for which
unitarity is violated in $FF\to VVV$ taking $f_{M,i}/\Lambda^4 =1$
TeV$^{-4}$:
\begin{equation}
  (14, 16, 14, 16, { 16}, 18, 19) \hbox{ TeV for }  i=0,\dots,7\;. 
  \label{eq:mtyp}
\end{equation}
For the same couplings unitarity is violated in $VV\to VV$ at diboson
center-of-mass energy $(2.5,3.6,2.4,3.4,3.2,3.8,4.2)$ TeV
respectively.

\subsection{ ${\cal O}_{T,i}$ operators}

All ${\cal O}_{T,i}$ operators possess four field strength tensors,
consequently their leading contributions appear in amplitudes where
all gauge bosons are transversely polarized.  Thus, in this case there
are no $v$ factors in the amplitude and, as expected from dimensional
analysis, the ${\cal O}_{T,i}$ operators lead to a growth of the
scattering amplitude proportional to $S_{VVV}^{3/2}$.  Hence, after
accounting for phase space and partial-wave $S_{VVV}$ dependence, we
find
\begin{equation}
 T^{1,a}_{-\frac{1}{2},\frac{1}{2};\pm 1,\pm 1, \pm 1}={\rm C}^{VVV}_{T,i}
 \times \frac{f_{T,i}}{\Lambda^4 }\, S_{VVV}^2\;.
\end{equation}

We present in Tables~\ref{tab:VVVft1} and ~\ref{tab:VVVft2} the
${\cal O}_{T,i}$ contributions to the non-orthogonal partial-wave
amplitudes for all the relevant processes.  After rotating to the
projections on the orthogonal basis with Eq.~\eqref{eq:rotation}, we
find the largest partial-wave amplitudes to have coefficients
\begin{equation}
  \left|
      {\rm C}^{VVV}_{T,i,\rm max}\right|  
      =\left( 1.9\,,\,2.0\,,\,0.61 \,,\, 0.74 \,,\,0.34 \,,\,0.94  \,,\,0.68\,,\,0.23\,,\,1.9 \,,\,0.55\right)\times 10^{-4}
 \label{eq:cvvvT} 
\end{equation}
for $i=0,1,2,3,4,5,6,7,8,9$, respectively. They arise from
$f_d\,\bar{f}_u \rightarrow W^- W^+ W^-$ for $i=0,1,2,3$,
$f_d\,\bar{f}_u \rightarrow \gamma\gamma W^-$ for $i=4,5,6,7$, and
$\ell \bar\ell\rightarrow \gamma\gamma\gamma$ and
$\nu_\ell \bar\nu_\ell\rightarrow \gamma\gamma\gamma$, for
$i=8,9$. This leads to the unitarity constraints
\begin{eqnarray}
&&  \left | \Frac{f_{T,0}}{\Lambda^4} \right|< 5.2 \times 10^{3} \Frac{\rm TeV^4}{S_{VVV}^2}\;,
  \hspace*{2cm} 
  \left | \Frac{f_{T,1}}{\Lambda^4} \right|< 5.0 \times 10^{3} \Frac{\rm TeV^4}{S_{VVV}^2}\;,\nonumber\\
&&  \left | \Frac{f_{T,2}}{\Lambda^4} \right|< 1.6 \times 10^{4} \Frac{\rm TeV^4}{S_{VVV}^2}\;,
  \hspace*{2cm} 
  \left | \Frac{f_{T,3}}{\Lambda^4} \right|< 1.3 \times 10^{4} \Frac{\rm TeV^4}{S_{VVV}^2}\;,\nonumber\\
&&  \left | \Frac{f_{T,4}}{\Lambda^4} \right|< 3.0 \times 10^{4} \Frac{\rm TeV^4}{S_{VVV}^2}\;,
  \hspace*{2cm} 
  \left | \Frac{f_{T,5}}{\Lambda^4} \right|< 1.1 \times 10^{4} \Frac{\rm TeV^4}{S_{VVV}^2}\;,\\
&&  \left | \Frac{f_{T,6}}{\Lambda^4} \right|< 1.5 \times 10^{4} \Frac{\rm TeV^4}{S_{VVV}^2}\;.
  \hspace*{2cm} 
  \left | \Frac{f_{T,7}}{\Lambda^4} \right|< 4.3 \times 10^{4} \Frac{\rm TeV^4}{S_{VVV}^2}\;,\nonumber\\
&&  \left | \Frac{f_{T,8}}{\Lambda^4} \right|< 5.2 \times 10^{3} \Frac{\rm TeV^4}{S_{VVV}^2}\;.
  \hspace*{2cm} 
  \left | \Frac{f_{T,9}}{\Lambda^4} \right|< 1.8 \times 10^{4} \Frac{\rm TeV^4}{S_{VVV}^2}\;.
\nonumber
\end{eqnarray}

So despite the ${\cal O}_{T,i}$ operators leading to a growth of the
scattering amplitude proportional to $S_{VVV}^2$, again, the
two-to-three unitarity limits are weaker than the ones from
$VV\rightarrow VV$. As for the case of ${\cal O}_{M,i}$ operators, the
corresponding numerical coefficients are smaller by 2-3 orders of
magnitude because of additional phase-space, partial-wave, and gauge
coupling factors, which result into looser unitarity bounds. For
instance, assuming $f_{T,i}/\Lambda^4 =1 $ TeV$^{-4}$, unitarity
violation takes place in $FF\to VVV$ at $\sqrt{S_{VVV}}$ of
\begin{equation}
  (8.5, 8.4, 11, 11, 13, 10, 11, 14, 8.5, 12)  \hbox{ TeV for }
  i=0,\dots, 9 \;,
\end{equation}
while in $VV\to VV$ unitarity violation occurs at
$(1.6,2.0,2.2,2.4,2.7,1.9,2.1,1.5,1.8.1.8)$ TeV, respectively.

\section{Discussion}
\label{sec:disc}

For a fixed value of a Wilson coefficient, the unitarity bound
indicates the energy scale where perturbative unitarity breaks down,
signaling the need to modify the theoretical approach at higher
energies. In particular, the analysis of the anomalous QGC
contributions to the $VV \to VV$ channel marks the energy limit of the
applicability of EFT as a function of the gauge boson pair invariant
mass. Analogously, the study of the $FF \to VVV$ process indicates the
maximum validity of the approach as a function of the
three-gauge-boson invariant mass.  From the previous section we have
learned that the QGC contributions to $FF \rightarrow VVV$ helicity
amplitudes grow as $S_{VVV}^2$ with $\sqrt{S_{VVV}}$ being the three
gauge boson invariant mass; see Table~\ref{tab:qgcs2}.
\smallskip 

\begin{table}
{\small \begin{tabular}{l|ccccccc|cccccccccc}
  Process & ${\cal O}_{M,0}$ &${\cal O}_{M,1}$ &${\cal O}_{M,2}$ &${\cal O}_{M,3}$ &${\cal O}_{M,4}$ &${\cal O}_{M,5}$
  &${\cal O}_{M,7}$ & ${\cal O}_{T,0}$ &${\cal O}_{T,1}$ &${\cal O}_{T,2}$ &${\cal O}_{T,3}$ &${\cal O}_{T,4}$ &${\cal O}_{T,5}$ &
  ${\cal O}_{T,6}$ &${\cal O}_{T,7}$ &${\cal O}_{T,8}$ & ${\cal O}_{T,9}$ \\\hline\hline
$f\bar{f} \rightarrow \gamma \gamma \gamma$ &    
  &  &  &  &  &  & &
  \checkmark
  & \checkmark &\checkmark  & \checkmark &\checkmark  &\checkmark  &\checkmark &\checkmark &\checkmark  & \checkmark    
 \\\hline
$f\bar{f} \rightarrow \gamma Z Z$ &
 \checkmark&\checkmark&\checkmark&\checkmark&\checkmark& \checkmark
  &\checkmark     &
 \checkmark & \checkmark &\checkmark  & \checkmark &\checkmark  &\checkmark
 &\checkmark &\checkmark &\checkmark  & \checkmark    
\\\hline
$f\bar{f} \rightarrow \gamma \gamma Z$ &
  &  &  &  &  &  & &
  \checkmark & \checkmark &\checkmark  & \checkmark &\checkmark  &\checkmark  &\checkmark &\checkmark &\checkmark  & \checkmark    
\\\hline
 $f\bar{f} \rightarrow  Z Z Z$ &
 \checkmark&\checkmark&\checkmark&\checkmark& \checkmark & \checkmark 
  &\checkmark     &
 \checkmark & \checkmark &\checkmark  & \checkmark & \checkmark &
 \checkmark& \checkmark& \checkmark &\checkmark  & \checkmark    
\\\hline
$f\bar{f} \rightarrow W^- W^+  \gamma$ &
 \checkmark&\checkmark&\checkmark&\checkmark&\checkmark& \checkmark
  &\checkmark     &
 \checkmark & \checkmark &\checkmark  & \checkmark &\checkmark  &\checkmark
 &\checkmark &\checkmark & &    
\\\hline
 $f\bar{f} \rightarrow W^- W^+ Z$ &
 \checkmark&\checkmark&\checkmark&\checkmark&\checkmark& \checkmark
  &\checkmark     &
 \checkmark & \checkmark &\checkmark  & \checkmark &\checkmark  &\checkmark
 &\checkmark &\checkmark & &    
\\\hline
 $f_d\,\bar{f}_u \rightarrow W^- W^+ W^-$ & 
 \checkmark&\checkmark&  & & & 
  &\checkmark     &
\checkmark & \checkmark &\checkmark  & \checkmark &  &
 & & & &    
\\\hline
$f_d\,\bar{f}_u \rightarrow \gamma \gamma W^-$ &
 &  &  & & &  &   &
 \checkmark & \checkmark &\checkmark  & \checkmark &\checkmark  &\checkmark
 &\checkmark &\checkmark & &    
\\\hline
$f_d\,\bar{f}_u \rightarrow  Z Z  W^-$&
 \checkmark&\checkmark&  & &\checkmark & \checkmark
 &\checkmark &
 \checkmark & \checkmark &\checkmark  & \checkmark &\checkmark  &\checkmark
 &\checkmark &\checkmark & &    
\\\hline
 $f_d\,\bar{f}_u \rightarrow \gamma Z W^- $ & 
  &  &  & & \checkmark &  \checkmark & \checkmark &
 \checkmark & \checkmark &\checkmark  & \checkmark &\checkmark  &\checkmark
 &\checkmark &\checkmark & &    
\end{tabular}\\
}
\caption{QGC contributions to $FF \to VVV$ that present a $S_{VVV}^2$
  growth of the helicity amplitude. }
\label{tab:qgcs2}
\end{table}

For a given center-of-mass energy, the unitarity bounds derived from
these amplitudes are always weaker than those from gauge boson pair
scattering.  This raises the question of how to implement the
unitarity constraints from gauge boson pair scattering when performing
searches for anomalous quartic couplings in triple gauge boson
production.  One conservative prescription, sometimes referred to as
``clipping method'' in the experimental
literature~\cite{ATLAS:2024nab, ATLAS:2025wyo}, consists in
restricting the EFT contribution to phase space regions where any
invariant mass of the gauge bosons pairs is smaller than a cutoff
value ($M_{VV,{\rm clip}}$). Next, the constraints on the Wilson
coefficient are derived from data as a function of
$M_{VV,{\rm clip}}$.  The limits are more stringent the higher
$M_{VV,{\rm clip}}$ is due to the growth of the anomalous cross
section and the validity of the limits are verified by comparing them
the corresponding unitarity bounds from $VV\rightarrow VV$. In some
cases of the LHC Run I and Run II data, the results show sensitivity
of the $FF\rightarrow VVV$ processes to QGC without violation of the
$VV\rightarrow VV$ unitarity bounds~\cite{ATLAS:2024nab,
  ATLAS:2025wyo} for $M_{VV,{\rm clip}}$ of the order 1-2 TeV.
\smallskip

Within this approach it is straightforward to clarify if the
$FF \to VVV$ unitarity limits have any impact on the QGC analyses of
three gauge boson production performed above.  This can be
systematically addressed by determining the maximum $S_{VVV}$ implied
by the $FF \to VVV$ unitarity limits for the limit value of the
coupling obtained after employing the clipping method. \smallskip

In brief, with the clipping method one obtains the bounds on the
anomalous coefficients from the three gauge boson process $P$
compatible with the $VV\rightarrow VV$ unitarity bounds to be
\begin{equation}
  \frac{f_X}{\Lambda^4} \leq \frac{1}{|{\rm C}^{VV}_{X,\rm{max}}|\,M^4_{VV{\rm clip},P}}
\label{eq:limvv}  
\end{equation}
where $|C^{VV}_{X,\rm{max}}|$ are the numerical coefficients in
Eq.~\eqref{eq:cvvmax}. Now, for this value of the anomalous Wilson
coefficient, $FF\rightarrow VVV$ unitarity requires that
\begin{equation}
\sqrt{S_{VVV,P}}\leq  \left(\frac{\Lambda^4}{f_X}\frac{1}{ |{\rm C}^{VVV}_{X,\rm{max}}|} \right)^{\frac{1}{4}} \;.
\label{eq:limvvvv}
\end{equation}
The most constraining case will correspond to the maximal value of the
Wilson coefficient in Eq.~\eqref{eq:limvv} which implies that the
maximum three-gauge boson invariant mass for which $FF\rightarrow VVV$
unitarity is not violated in process $P$ is
\begin{equation} \sqrt{S^{{\rm max}}_{VVV,P}} =
  \left(\frac{|C^{VV}_{X,{\rm max}}|}{|C^{VVV}_{X,{\rm max}}|}\right)^{\frac{1}{4}}\,
       M_{VV{\rm clip},P} \sim 4-6 \,M_{VV{\rm clip},P}\;,
\end{equation}
where we used the values of the coefficients
${\rm C}^{VV}_{X,{\rm max}}$ in Eq.~\eqref{eq:cvvmax} and the
coefficients ${\rm C}^{VVV}_{X,{\rm max}}$ in Eqs.~\eqref{eq:cvvvM}
and ~\eqref{eq:cvvvT}. \smallskip

On the other hand, simple kinematics implies that for any process $P$
with the tree final gauge bosons with four-momenta $p_1$, $p_2$ and
$p_3$,
\begin{eqnarray}
  \sqrt{S_{VVV,P}}
  &\equiv&\sqrt{(p_1+p_1+p_3)^2}\nonumber \\
  &\leq&
    \sqrt{(p_1+p_2)^2+(p_1+p_3)^2+(p_2+p_3)^2}=
    \sqrt{M_{12}^2+M_{13}^2+M_{23}^2}\,\leq\,
  \sqrt{3}\, M_{VV{\rm clip},P}\, ,
\label{eq:svvv}
\end{eqnarray}
where $M_{ij}$ stands for the invariant mass of the particles $i$ and
$j$. So $\sqrt{S_{VVV,P}}$ is always smaller than
$\sqrt{S^{{\rm max}}_{VVV,P}}$ by at least a factor 2.  \smallskip

All in all we conclude that the $FF \to VVV$ unitarity bounds do not
alter the present~\cite{ATLAS:2017bon, ATLAS:2022xnu, ATLAS:2022wmu,
  ATLAS:2024nab, ATLAS:2025wyo, CMS:2014cdf, CMS:2021jji, CMS:2023rcv,
  CMS:2025oey} and future anomalous QGC analyses in the exclusive
production of three gauge bosons obtained with the clipping method. In
brief, our findings demonstrate the consistency of the clipping
method.  \smallskip

\acknowledgments

We thank Gabriel M. Salla for enlightening discussions.  OJPE is
partially supported by CNPq grant number 302120/2025-4.  This project
is funded by USA-NSF grant PHY-2210533.  It has also received support
from the European Union's Horizon 2020 research and innovation program
under the Marie Sk\l odowska-Curie grant agreement No 860881-HIDDeN,
and Horizon Europe research and innovation programme under the Marie
Sk\l odowska-Curie Staff Exchange grant agreement No 101086085 --
ASYMMETRY''.  It also receives support from grants
PID2019-105614GB-C21, and ``Unit of Excellence Maria de Maeztu
2020-2023'' award to the ICC-UB CEX2019-000918-M, funded by
MCIN/AEI/10.13039/501100011033, and from grant 2021-SGR-249
(Generalitat de Catalunya).

\newpage
\appendix

\section{Helicity amplitudes}
\label{app:hel}

Here, we present the helicity amplitudes in
Tables~\ref{tab:000}--\ref{tab:VVVft2}
that contain the fastest growth with $\sqrt{S}$ for the processes
$FF \to VVV$ ($V=\gamma$, $Z$, $W^\pm$, and $H$) used to obtain the
$2 \to 3$ unitarity constraints.\smallskip

\begin{table}[h]
\begin{tabular}{|l|c|c@{\extracolsep{1cm}}c@{\extracolsep{0.08cm}}
c|}\hline
  &    &\multicolumn{3}{c|}
  {$\tilde{T}^{1,a}_{-\frac{1}{2},\frac{1}{2};0,0,0}$
in units of 
$ \frac{e^2}{3840 \pi^2}\times g_L^f \times \frac{f_{S,i} }{\Lambda^4} \, S_{VVV}^{3/2}\, v
\times \frac{1}{\sqrt{N_{id}!}}$}\\\hline
  & a/i  & 0 & 1 & 2 
  \\\hline
  $f\,\bar{f}
  \rightarrow W^+W^- Z$
    & 1 & $-i\frac{1}{2\sqrt{2} c_W^2 s_W^2}$
    &  $i\frac{1}{\sqrt{2} c_W^2 s_W^2}$ &
  $-i\frac{1}{2\sqrt{2} c_W^2 s_W^2}$
    \\
    & 2 & $i\frac{1}{\sqrt{2} c_W^2 s_W^2}$
    & $-i\frac{\sqrt{2}}{ c_W^2 s_W^2}$ &
    $i\frac{1}{\sqrt{2} c_W^2 s_W^2}$
    \\\hline     
    $f\,\bar{f}
    \rightarrow ZHH$
    & 1
    & $-i\frac{\sqrt{2}}{ c_W^2 s_W^2}$
    & $i\frac{1}{\sqrt{2} c_W^2 s_W^2}$
    & $i\frac{1}{\sqrt{2} c_W^2 s_W^2}$
    \\
    & 2
    & $-i \frac{\sqrt{2}}{ c_W^2 s_W^2}$
    & $i\frac{1}{\sqrt{2} c_W^2 s_W^2}$
    & $i\frac{1}{\sqrt{2} c_W^2 s_W^2}$
   \\\hline      
   $f_d\,\bar{f}_u
   \rightarrow W^- W^+ W^-$
    & 1
   & $i\frac{1}{ s_W^2}$
   & $-i\frac{1}{2 s_W^2}$
   & $-i\frac{1}{2 s_W^2}$
   \\
    & 2
    & $i\frac{1}{ s_W^2}$
    &$-i\frac{1}{2 s_W^2}$
   &$-i\frac{1}{2 s_W^2}$
   \\\hline      
$f_d\,\bar{f}_u\rightarrow ZZ W^-$
   & 1
   & $-i\frac{1}{4 s_W^2}$
   & $i\frac{1}{2 s_W^2}$
   & $-i\frac{1}{4 s_W^2}$
   \\
   & 2
   & $i\frac{1}{2 s_W^2}$
   & $-i\frac{1}{ s_W^2}$
   & $i\frac{1}{2 s_W^2}$ 
   \\\hline      
   $f_d\,\bar{f}_u
   \rightarrow HH W^-$
   & 1
   & $-i\frac{1}{4 s_W^2}$
   & $i\frac{1}{2 s_W^2}$
   & $-i\frac{1}{4 s_W^2}$
   \\
   & 2
   & $i\frac{1}{2 s_W^2}$
   & $-i\frac{1}{ s_W^2}$
   & $i\frac{1}{2 s_W^2}$ 
   \\\hline      
$f_d\,\bar{f}_u\rightarrow ZH W^-$
   & 1
   & $\frac{3}{4 s_W^2}$
   & 0 
   & $-\frac{3}{4 s_W^2}$ 
   \\
   & 2
   & 0
   & 0
   & 0 \\\hline
\end{tabular}
\caption{Partial-wave non-orthogonal projections for $FF\to VVV$
  generated by ${\cal O}_{S,i}$.  In this Table, $f=u,d,\ell$, or
  $\nu_\ell$, while we denote by $f_{u(d)}=u (d)$ or
  $\nu_\ell(\ell)$. The neutral current (NC) scattering amplitude
  dependence on the chosen fermion is encoded in the couplings
  $g_L=T_3- Q s_W^2$ where $T_3$ and $Q$ are the third component of
  isospin and charge of the corresponding fermion with $s_W$ ($c_W$)
  being the sine (cosine) of the weak mixing angle. Moreover, $g_L=1$
  for the remaining charged current (CC) processes where we omit, for
  simplicity, the CKM factors. }
\label{tab:000}
\end{table}

\begin{table}[H]
\centering
{\small
\begin{tabular}{|l|c|ccccccc|}\hline
&     &\multicolumn{7}{c|}
  {$\tilde{T}^{1,a}_{-\frac{1}{2},\frac{1}{2};\lambda_3,\lambda_4,\lambda_5}$
    in units of 
    $\frac{e}{\pi^2} \frac{1}{192 \sqrt{70} }  \times \frac{f_{M,i}}
    {\Lambda^4} \, S_{VVV}^2 \times \frac{1}{\sqrt{N_{id}!}}$}\\\hline
  & a/i  & 0 & 1 & 2 & 3 &4 & 5 &7\\\hline
  $\lambda_3,\lambda_4,\lambda_5=0,0,1$
  &    &   &   &   &   &  &   & \\\hline
  $f\,\bar{f}\rightarrow Z Z \gamma $
  & 1 & $- T_3$ & $ \frac{2T_3}{3}$ & $Y$ & $ -\frac{2Y}{3}$ &
  $-\frac{\tilde{g}}{2 c_W s_W}$ &$ \frac{\tilde{g}  }{3 c_W s_W}$ &
  $ \frac{T_3}{3}$\\
  & 2 & $- T_3$ & $ \frac{2T_3}{3}$ & $Y$ & $ -\frac{2Y}{3}$ &
  $-\frac{\tilde{g}}{2 c_W s_W}$ &$ \frac{\tilde{g}}{3 c_W s_W}$ &
  $ \frac{2T_3}{6}$\\\hline
  $f\,\bar{f}\rightarrow Z Z Z$
  & 1 & $ -\frac{T_3c_W}{ s_W}$ & $ \frac{2T_3c_W}{3 s_W}$ & $  \frac{Ys_W}{c_W}$ & $ -\frac{2 Y s_W}{3 c_W}$ &
  $-\frac{2T_3-Y}{4}$  & $-\frac{2T_3-Y}{6}$
  & $ -\frac{2T_3 c_W }{6 s_W}$\\ 
  & 2 &$ -\frac{T_3c_W}{ s_W}$ & $ \frac{2T_3c_W}{3 s_W}$ & $  \frac{Ys_W}{c_W}$ & $ -\frac{2 Y s_W}{3 c_W}$ &
  $-\frac{2T_3-Y}{4}$ & $-\frac{2T_3-Y}{6}$&
  $ -\frac{2T_3 c_W }{6 s_W}$\\ \hline
  $f\,\bar{f}\rightarrow W^+ W^- \gamma$
  & 1 &  $ -T_3$ & $ \frac{2T_3}{3}$ & $-Y$ & $ \frac{2Y}{3}$ &
  $ -\frac{\tilde{g}}{3 c_W s_W}$
  & $- \frac{10 T_3 c_W^2 +3 Ys_W^2}{24 c_W s_W}$ & $ -\frac{T_3}{3}$\\ 
  & 2 &   $ -T_3$ & $ \frac{2T_3}{3}$ & $-Y$ & $ \frac{2Y}{3}$ &
  $ -\frac{\tilde{g}}{2 c_W s_W}$
  & $- \frac{6 T_3 c_W^2 +5 Ys_W^2}{24 c_W s_W}$ & $ -\frac{T_3}{3}$\\ \hline
  $f\,\bar{f}\rightarrow  W^+ W^- Z  $
  & 1 & $ -\frac{T_3c_W}{ s_W}$ & $ \frac{2T_3c_W}{3s_W}$ & $ \frac{Ys_W}{c_W}$ & $ -\frac{2Y  s_W}{3 c_W}$ &
  $\frac{2T_3-Y}{4}$ & $ -\frac{-10 T_3  +3 Y}{24}$ & $ -\frac{T_3 c_W }{3 s_W}$\\ 
  & 2 &  $ -\frac{T_3c_W}{ s_W}$ & $ \frac{2T_3c_W}{3s_W}$ & $ \frac{Ys_W}{c_W}$ & $ -\frac{2Y  s_W}{3 c_W}$ &
  $\frac{2T_3-Y}{4}$ & $ -\frac{-10 T_3  +3 Y}{24}$ & $ -\frac{T_3 c_W }{3 s_W}$\\ \hline
  $f\,\bar{f}\rightarrow  H H   \gamma$
  &1 & $ -T_3$ & $ \frac{2T_3}{3}$ & $-Y$ & $ \frac{2Y}{3}$ &
  $  \frac{\tilde{g}}{2 c_W s_W}$ & $ \frac{\tilde{g}}{3 c_W s_W}$ & $ -\frac{T_3}{3}$\\
  &2 & $ -\frac{2T_3}{2}$ & $ \frac{2T_3}{3}$ & $-Y$ & $ \frac{2Y}{3}$ &
  $  \frac{\tilde{g}}{2 c_W s_W}$ & $ \frac{\tilde{g}}{3 c_W s_W}$ & $ -\frac{T_3}{3}$\\  \hline
  $f\,\bar{f}\rightarrow  H H Z $
  & 1 & $  -\frac{T_3c_W }{s_W}$ & $ \frac{2T_3 c_W }{3 s_W}$ & $ \frac{Ys_W}{c_W}$ & $ -\frac{2Y  s_W}{3 c_W}$ &
  $-\frac{2T_3-Y}{4}$ & $-\frac{2T_3-Y}{6}$& $  -\frac{T_3c_W }{3 s_W}$\\
  & 2 &$  -\frac{T_3c_W }{ s_W}$ & $ \frac{2T_3c_W }{3 s_W}$ & $ \frac{Ys_W}{c_W}$ & $ -\frac{2Y  s_W}{3 c_W}$ &
  $-\frac{2T_3-Y}{4}$ & $-\frac{2T_3-Y}{6}$ & $  -\frac{T_3c_W }{3 s_W}$\\
  \hline
  $f\,\bar{f}\rightarrow H W^+ W^- $
  &1 &  $ 0$ & $ 0$ & $ 0$ & $ 0$ & $  -i\frac{Y}{4 c_W}$ & $ -i \frac{5Y}{24 c_W}$
  & $ i\frac{2T_3}{24 s_W}$\\ 
  &2 &   $ 0$ & $ 0$ & $ 0$ & $ 0$ & $  -i\frac{Y}{4 c_W}$ & $ -i \frac{3Y}{24 c_W}$
  & $ i\frac{2T_3}{24 s_W}$\\ 
  \hline
  $f_d\,\bar{f}_u
  \rightarrow  W^-  W^+  W^-$
  & 1 & $ -\frac{1}{\sqrt{2} s_W}$ & $  \frac{\sqrt{2}}{3 s_W}$ & $ 0$ & $ 0$ & $ 0$ & $ 0$ &
  $ -\frac{1}{ 4 \sqrt{2} s_W}$\\
  & 2 & $ -\frac{1}{\sqrt{2} s_W}$ & $  \frac{\sqrt{2}}{3 s_W}$ & $ 0$ & $ 0$ & $ 0$ & $ 0$ & $ -\frac{5}{12 \sqrt{2} s_W}$\\  \hline
  $f_d\,\bar{f}_u\rightarrow  Z Z  W^-$
  & 1 & $ -\frac{1}{\sqrt{2} s_W}$ & $  \frac{\sqrt{2}}{3 s_W}$ & $ 0$ & $ 0$ & $ 0$ & $ 0$ & $ -\frac{1}{3 \sqrt{2} s_W}$\\
  & 2 & $ -\frac{1}{\sqrt{2} s_W}$ & $  \frac{\sqrt{2}}{3 s_W}$ & $ 0$ & $ 0$ & $ 0$ & $ 0$ & $ -\frac{1}{3 \sqrt{2} s_W}$\\  \hline
  $f_d\,\bar{f}_u\rightarrow Z W^- \gamma $
  &1 &  $ 0$ & $ 0$ & $ 0$ & $ 0$ & $ \frac{c_W }{2 \sqrt{2} s_W}$ & $ \frac{5 c_W }{12 \sqrt{2} s_W}$ & $ \frac{1}{12 \sqrt{2}}$\\
  &2 &  $ 0$ & $ 0$ & $ 0$ & $ 0$ & $ \frac{c_W }{2 \sqrt{2} s_W}$ & $ \frac{ c_W }{4 \sqrt{2} s_W}$ & $ -\frac{1}{12 \sqrt{2}}$\\\hline 
  $f_d\,\bar{f}_u\rightarrow H H W^-$
  &1 & $ -\frac{1}{\sqrt{2} s_W}$ & $ \frac{\sqrt{2} }{3 s_W}$ & $ 0$ & $ 0$ & $ 0$ & $ 0$ & $ -\frac{1}{3 \sqrt{2} s_W}$\\
  &2 & $ -\frac{1}{\sqrt{2} s_W}$ & $ \frac{\sqrt{2} }{3 s_W}$ & $ 0$ & $ 0$ & $ 0$ & $ 0$ & $ -\frac{1}{3 \sqrt{2} s_W}$\\\hline  
  $f_d\,\bar{f}_u\rightarrow H W^- \gamma $
  &1 &  $  0$ & $ 0$ & $ 0$ & $ 0$ & $i \frac{ c_W }{2 \sqrt{2} s_W}$ & $ i\frac{5 c_W }{12 \sqrt{2} s_W}$ & $  i\frac{1}{12 \sqrt{2}}$\\
  &2 &  $  0$ & $ 0$ & $ 0$ & $ 0$ & $i \frac{ c_W }{2 \sqrt{2} s_W}$ & $ i\frac{ c_W }{4 \sqrt{2} s_W}$ & $  -i\frac{1}{12 \sqrt{2}}$\\\hline
  $f_d\,\bar{f}_u\rightarrow H W^-  Z$
  & 1 & $ 0$ & $ 0$ & $ 0$ & $ 0$ & $  -i\frac{1}{2 \sqrt{2}}$ & $ -i\frac{5 }{12 \sqrt{2}}$ & $  i \frac{c_W }{12 \sqrt{2} s_W}$\\
  & 2 & $ 0$ & $ 0$ & $ 0$ & $ 0$ & $  -i\frac{1}{2 \sqrt{2}}$ & $ -i\frac{1 }{4 \sqrt{2}}$ & $  -i \frac{c_W }{12 \sqrt{2} s_W}$\\
  \hline
   $\lambda_3,\lambda_4,\lambda_5=0,1,0$ 
  &    &   &   &   &   &  &   &  \\\hline 
$f\bar{f} \rightarrow  W^+ W^- Z  $ & 1
  & $ 0$ & $ 0$ & $ 0$ & $ 0$ & $ \frac{Y}{4 c_W}$ & $ \frac{Y}{8 c_W}$ & $ \frac{T_3}{12 s_W}$ \\
 & 2
 & $ 0$ & $ 0$ & $ 0$ & $ 0$ & $ \frac{Y}{ 4 c_W}$ & $ \frac{5 Y}{24 c_W}$ & $ -\frac{T_3}{12 s_W}$ \\\hline
$f\bar{f} \rightarrow H W^+W^- $ & 1
& $ 0$ & $ 0$ & $ 0$ & $ 0$ & $  i\frac{Y}{4 c_W}$ & $ i\frac{Y}{8 c_W}$ & $ -i\frac{T_3}{12 s_W}$ \\
 &2
 & $ 0$ & $ 0$ & $ 0$ & $ 0$ & $ i\frac{Y}{4 c_W}$ & $ i\frac{5 Y}{24 c_W}$ & $ i\frac{T_3}{12 s_W}$ \\
  \hline
$f_d\,\bar{f}_u \rightarrow  Z Z  W^-$ & 1 &  $ 0$ & $ 0$ & $ 0$ & $ 0$   
&  $-\frac{1}{2\sqrt{2}}$ & $ -\frac{5}{12\sqrt{2}}$ & $ \frac{c_W}{12\sqrt{2}s_W}$ \\ 
 & 2 &  $ 0$ & $ 0$ & $ 0$ & $ 0$   
 &  $-\frac{1}{2\sqrt{2}}$ & $ -\frac{1}{4\sqrt{2}}$ & $ -\frac{c_W}{12\sqrt{2}s_W}$ \\  \hline
$f_d\,\bar{f}_u \rightarrow H W^-  Z$ & 1
&$ 0$ & $ 0$ & $ 0$ & $ 0$ & $ 0$ & $ 0$ & $ -i\frac{1}{12 s_W}$ \\
& 2
&$ 0$ & $ 0$ & $ 0$ & $ 0$ & $ 0$ & $ 0$ & $ i\frac{1}{12 s_W}$ \\
\hline
\end{tabular}}
\caption{Partial wave non-orthogonal projections for
  $FF \rightarrow VVV$ generated by ${\cal O}_{M,i}$.  In the table
  $f=u,d,\ell$, or $\nu_\ell$, while we denote by $f_{u(d)}=u (d)$ or
  $\nu_\ell(\ell)$. The scattering amplitude dependence on the chosen
  fermion is encoded in the couplings
  $g_L=T_{3}- Q s_W^2=T_3c_W^2-\frac{Y}{2}s_W^2$ with the hypercharge
  $Y=2(Q-T_3)$, and the we have defined the combination
  $\tilde{g}\equiv T_3 c_W^2+\frac{Y}{2}s_W^2$. The amplitude for the
  CC processes are the same for quarks and leptons up to the CKM
  factor which we are omitting for simplicity.}

\label{tab:VVVfm}
\end{table}

\begin{table}[H]
\scalebox{0.80}{
\begin{tabular}{|l|cccccccccc|}\hline
    &\multicolumn{10}{c|}
  {$\tilde{T}^{1,a}_{-\hf,\hf;\lambda_3,\lambda_4,\lambda_5}=\tilde{T}^{1,2}_{-\hf,\hf;\lambda_3,\lambda_4,\lambda_5}$
    in units of 
    $\frac{e}{\pi^2} \sqrt{\frac{7}{155}}
    \times \frac{f_{T,i}}{\Lambda^4} \, S_{VVV}^2
\times \frac{1}{\sqrt{N_{id}!}}$}\\\hline
             & 0 & 1 & 2 & 3 &4 & 5 &6 &7 &8 &9\\\hline
$\lambda_3,\lambda_4,\lambda_5=1,1,-1$
  &\multicolumn{10}{c|}{}   \\\hline
$f\bar{f} \rightarrow \gamma \gamma \gamma$ &    
  $-\frac{ T_3s_W^2}{48 \sqrt{2}}$ &  
  $-\frac{ T_3 s_W^2}{48 \sqrt{2}}$ &
  $-\frac{ T_3s_W^2}{64 \sqrt{2}}$ &
  $-\frac{ T_3s_W^2}{64 \sqrt{2}}$ &
  $-\frac{\tilde{g}}{64 \sqrt{2}}$ &
  $-\frac{\tilde{g}}{48 \sqrt{2}}$ &
$-\frac{\tilde{g}}{48 \sqrt{2}}$ &    
$-\frac{\tilde{g}}{48 \sqrt{2}}$ &
  $-\frac{Yc_W^2 }{24 \sqrt{2}}$ &
  $-\frac{Y c_W^2 }{32 \sqrt{2}}$ 
\\
$f \bar{f} \rightarrow \gamma Z Z$ &
$-\frac{ T_3 c_W^2 }{48 \sqrt{2}}$ &  
$-\frac{ T_3 c_W^2 }{48 \sqrt{2}}$ &
$-\frac{ T_3 c_W^2 }{64 \sqrt{2}}$ &
$-\frac{  T_3c_W^2 }{64 \sqrt{2}}$ &
$-\frac{Q-3\tilde{g}}{192 \sqrt{2}}$ &
$\frac{\tilde{g}}{48\sqrt{2}}$ &  
${{\frac{(c_W^2-s_W^2)(2T_3-Y)}{192\sqrt{2}}}}$  &  
$-\frac{Q-3\tilde{g}}{192 \sqrt{2}}$ &
$-\frac{Y s_W^2}{24 \sqrt{2}}$ &
$- \frac{Y s_W^2}{32 \sqrt{2}}$ 
\\
$f \bar{f} \rightarrow \gamma \gamma Z$ &
$-\frac{ T_3c_W s_W}{48 \sqrt{2}}$ &  
 $-\frac{ T_3c_W  s_W}{48 \sqrt{2}}$ &
$-\frac{ T_3 c_W s_W}{64 \sqrt{2}}$ &
$-\frac{ T_3c_W s_W}{64 \sqrt{2}}$ &
$-\frac{(1-3s_W^2)(g_L)}{192 \sqrt{2} c_W s_W}$&
$\frac{-2 c_W^4 T_3+s_W^4 Y}{96 \sqrt{2} c_W s_W}$&
${{\frac{(2T_3-Y)s_W c_W}{96 \sqrt{2}}}}$  &  
$-\frac{(1-3s_W^2)(g_L)}{192 \sqrt{2} c_W s_W}$&
$\frac{Y c_W s_W}{24\sqrt{2}}$ &
$\frac{Y c_W s_W}{32\sqrt{2}}$ 
\\
 $f\bar{f} \rightarrow  Z Z Z$ &
 $-\frac{ T_3 c_W^3}{48 \sqrt{2} s_W}$ &  
 $-\frac{ T_3 c_W^3}{48 \sqrt{2} s_W}$ &
$-\frac{ T_3c_W^3}{64 \sqrt{2} s_W}$ &
$-\frac{ T_3 c_W^3}{64 \sqrt{2} s_W}$ &
${{-\frac{(2T_3-Y)c_W s_W}{128\sqrt{2}}}}$  &
${{-\frac{(2T_3-Y)c_W s_W}{96\sqrt{2}}}}$  &
${{-\frac{(2T_3-Y)c_W s_W}{96\sqrt{2}}}}$  &
${{-\frac{(2T_3-Y)c_W s_W}{128\sqrt{2}}}}$  & 
$\frac{Y s_W^3}{24\sqrt{2} c_W}$ &
$\frac{Y s_W^3}{32\sqrt{2} c_W}$  
\\
 $f\bar{f} \rightarrow W^- W^+  \gamma$ &
$-\frac{ T_3}{48 \sqrt{2}}$ &
0 &
$-\frac{ T_3}{192 \sqrt{2}}$ &
$-\frac{ T_3}{192 \sqrt{2}}$ &
$-\frac{Y}{ 384 \sqrt{2}}$ &
$-\frac{Y}{ 96 \sqrt{2}}$ &
0 &  
$-\frac{Y}{384 \sqrt{2}}$ &
0 &
0
  \\
 $f\bar{f} \rightarrow W^- W^+ Z$ &
  $-\frac{ T_3c_W}{48 \sqrt{2} s_W}$&
0 &
  $-\frac{ T_3 c_W }{192 \sqrt{2} s_W}$ &
  $-\frac{ T_3c_W }{192 \sqrt{2} s_W}$ &
  $\frac{Y c_W }{384 \sqrt{2} s_W}$ &  
$ \frac{Y s_W}{96\sqrt{2} c_W}$ &
0 &
$ \frac{Ys_W}{384\sqrt{2} c_W}$ &
0 &
0
\\
 $f_d\,\bar{f}_u \rightarrow W^- W^+ W^-$ & 
 $-\frac{1}{96 s_W}$ &
 $-\frac{1}{96 s_W}$ &  
$-\frac{1}{192 s_W}$ &  
$-\frac{1}{192 s_W}$ &  0 &  0 &  0 &  0 &  0 &  0
\\
 $f_d\,\bar{f}_u\rightarrow \gamma \gamma W^-$ & 
$-\frac{s_W}{96}$ &
0 &  
$-\frac{s_W}{384}$ &
$-\frac{s_W}{384}$ &
$-\frac{c^2_W}{384 s_W}$ &
$-\frac{c^2_W}{95 s_W}$ &
0 &
$-\frac{c_W^2 }{384 s_W}$ &  
  0 &  0
\\
$f_d\,\bar{f}_u\rightarrow  Z Z  W^-$&
$-\frac{c_W^2}{96 s_W}$ &  0 &  
$-\frac{c_W^2}{384 s_W}$ &
$-\frac{c_W^2}{384 s_W}$ &
$-\frac{s_W}{384}$ &
$-\frac{s_W}{96}$ &  0 &
$-\frac{s_W}{384}$ &
0 &  0
\\
$f_d\,\bar{f}_u \rightarrow \gamma Z W^- $ & 
 $-\frac{c_W}{96}$ &  0 &
$-\frac{c_W}{384}$ &
$-\frac{c_W}{384}$ &
$\frac{c_W}{384}$ &
 $ \frac{c_W}{96}$ &  0 &
 $ \frac{c_W}{384}$ &  0 &  0
\\
\hline
$\lambda_3,\lambda_4,\lambda_5=1,-1,1$
&\multicolumn{10}{c|}{}   \\\hline
$f\bar{f} \rightarrow \gamma \gamma \gamma$ &    
  $-\frac{ T_3s_W^2}{48 \sqrt{2}}$ &  
  $-\frac{ T_3 s_W^2}{48 \sqrt{2}}$ &
  $-\frac{ T_3s_W^2}{64 \sqrt{2}}$ &
  $-\frac{ T_3s_W^2}{64 \sqrt{2}}$ &
  $-\frac{\tilde{g}}{64 \sqrt{2}}$ &
  $-\frac{\tilde{g}}{48 \sqrt{2}}$ &
$-\frac{\tilde{g}}{48 \sqrt{2}}$ &    
$-\frac{\tilde{g}}{64 \sqrt{2}}$ &
  $-\frac{Yc_W^2 }{24 \sqrt{2}}$ &
  $-\frac{Y c_W^2 }{32 \sqrt{2}}$ 
\\
$f \bar{f} \rightarrow \gamma Z Z$ &
$-\frac{ T_3 c_W^2 }{48 \sqrt{2}}$ &  
$-\frac{ T_3 c_W^2 }{48 \sqrt{2}}$ &
$-\frac{ T_3 c_W^2 }{64 \sqrt{2}}$ &
$-\frac{  T_3c_W^2 }{64 \sqrt{2}}$ &
$-\frac{ Q-3\tilde{g}}{192 \sqrt{2}}$ &
$\frac{\tilde{g}}{48\sqrt{2}}$ &  
${{\frac{(c_W^2-s_W^2)(2T_3-Y)}{192\sqrt{2}}}}$  &  
$-\frac{ Q-3\tilde{g}}{192 \sqrt{2}}$ &
$-\frac{Y s_W^2}{24 \sqrt{2}}$ &
$- \frac{Y s_W^2}{32 \sqrt{2}}$ 
\\
$f \bar{f} \rightarrow \gamma \gamma Z$ &
$-\frac{ T_3c_W s_W}{48 \sqrt{2}}$ &  
 $-\frac{2 T_3c_W  s_W}{48 \sqrt{2}}$ &
$-\frac{ T_3 c_W s_W}{64 \sqrt{2}}$ &
$-\frac{ T_3c_W s_W}{64 \sqrt{2}}$ &
$-\frac{(1-3s_W^2)(g_L)}{192 \sqrt{2} c_W s_W}$&
${{\frac{(2T_3-Y)s_W c_W}{96 \sqrt{2}}}}$  &  
$-\frac{(\tilde{g})(c_W^2-s_W^2}{96 \sqrt{2} c_W s_W}$&
$-\frac{(1-3s_W^2)( g_L)}{192 \sqrt{2} c_W s_W}$&
$\frac{Y c_W s_W}{24\sqrt{2}}$ &
$\frac{Y c_W s_W}{32\sqrt{2}}$ 
\\
 $f\bar{f} \rightarrow  Z Z Z$ &
$-\frac{ T_3c_W^3}{48 \sqrt{2} s_W}$ &  
 $-\frac{2 T_3 c_W^3}{48 \sqrt{2} s_W}$ &
$-\frac{ T_3c_W^3}{64 \sqrt{2} s_W}$ &
$-\frac{ T_3 c_W^3}{64 \sqrt{2} s_W}$ &
${{-\frac{(2T_3-Y)c_W s_W}{128\sqrt{2}}}}$  &
${{-\frac{(2T_3-Y)c_W s_W}{96\sqrt{2}}}}$  &
${{-\frac{(2T_3-Y)c_W s_W}{96\sqrt{2}}}}$  &
${{-\frac{(2T_3-Y)c_W s_W}{128\sqrt{2}}}}$  & 
$\frac{Y s_W^3}{24\sqrt{2} c_W}$ &
$\frac{Y s_W^3}{32\sqrt{2} c_W}$  
\\
$f\bar{f} \rightarrow W^- W^+  \gamma$ &
$0$ &
$-\frac{ T_3}{96 \sqrt{2}}$ &
$ -\frac{ T_3}{192 \sqrt{2}}$ &
$ - \frac{ T_3}{192 \sqrt{2}}$ &
$ -\frac{Y}{384 \sqrt{2}}$ &
$ 0$ &
$-\frac{Y}{192 \sqrt{2}}$ &
$ - \frac{Y}{384 \sqrt{2}}$ &
$ 0$ & $ 0$
\\
 $f\bar{f} \rightarrow W^- W^+ Z$ &
$0$ &
 $ -\frac{ T_3c_W}{96 \sqrt{2} s_W}$ &
$  -\frac{ T_3c_W }{192 \sqrt{2} s_W}$ &
$  -\frac{ T_3c_W }{192 \sqrt{2} s_W}$ &
$ \frac{Ys_W}{384\sqrt{2} c_W}$ &
$ 0$ &
$ \frac{Y s_W}{192\sqrt{2} c_W}$ &
$ \frac{Y s_W}{384\sqrt{2} c_W}$ &
$ 0$ &$0$
\\
$f_d\,\bar{f}_u \rightarrow W^- W^+ W^-$ & 
 $ -\frac{1}{96 s_W}$ &
 $ -\frac{1}{192 s_W}$ &
$ -\frac{1}{192 s_W}$ & 
$ -\frac{1}{192 s_W}$ & 
$ 0$ & $ 0$ & $ 0$ & $ 0$ & $ 0$ & $ 0$
\\
$f_d\,\bar{f}_u\rightarrow \gamma \gamma W^-$
& $ 0$ & 
 $ -\frac{s_W}{192}$ &
$ -\frac{s_W}{384}$ & 
$ -\frac{s_W}{384}$ & 
$ -\frac{c_W^2}{384 s_W}$ &
$ 0$ &
$ -\frac{c_W^2}{192 s_W}$ &
$ -\frac{c_W^2}{384 s_W}$ &
$ 0$ & $ 0$
\\
$f_d\,\bar{f}_u\rightarrow  Z Z  W^-$&
 $ 0$ &
 $  -\frac{c_W^2 }{192 s_W}$ &
$ - \frac{c_W^2}{384 s_W}$ & 
$ - \frac{c_W^2}{384 s_W}$ &
$ - \frac{s_W}{384}$ &
$ 0$ &
$ - \frac{s_W}{192}$ &
$ - \frac{s_W}{384}$ &
$ 0$ & $ 0$
\\
$f_d\,\bar{f}_u \rightarrow \gamma Z W^- $ & 
$ 0$ &
$ - \frac{c_W}{192}$ &
$ - \frac{c_W}{384}$ &
$ - \frac{c_W}{384}$ &
$\frac{c_W}{384}$ &
$ 0$ &
$ \frac{c_W}{192}$ &
$\frac{c_W}{384}$ &
$ 0$ & $ 0$
\\
\hline
$\lambda_3,\lambda_4,\lambda_5=-1,1,1$ &\multicolumn{10}{c|}{}   \\\hline
$f\bar{f} \rightarrow \gamma \gamma \gamma$ &    
  $-\frac{ T_3s_W^2}{48 \sqrt{2}}$ &  
  $-\frac{ T_3 s_W^2}{48 \sqrt{2}}$ &
  $-\frac{ T_3s_W^2}{64 \sqrt{2}}$ &
  $-\frac{ T_3s_W^2}{64 \sqrt{2}}$ &
  $-\frac{\tilde{g}}{64 \sqrt{2}}$ &
  $-\frac{\tilde{g}}{48 \sqrt{2}}$ &
$-\frac{\tilde{g}}{48 \sqrt{2}}$ &    
$-\frac{\tilde{g}}{64 \sqrt{2}}$ &
  $-\frac{Yc_W^2 }{24 \sqrt{2}}$ &
  $-\frac{Y c_W^2 }{32 \sqrt{2}}$ 
\\
$f \bar{f} \rightarrow \gamma Z Z$ &
$-\frac{ T_3 c_W^2 }{48 \sqrt{2}}$ &  
$-\frac{ T_3 c_W^2 }{48 \sqrt{2}}$ &
$-\frac{ T_3 c_W^2 }{64 \sqrt{2}}$ &
$-\frac{ T_3c_W^2 }{64 \sqrt{2}}$ &
$-\frac{Q-3\tilde{g}}{192 \sqrt{2}}$ &
$\frac{\tilde{g}}{48\sqrt{2}}$ &  
${{\frac{(c_W^2-s_W^2)(2T_3-Y)}{192\sqrt{2}}}}$  &  
$-\frac{ Q-3\tilde{g}}{192 \sqrt{2}}$ &
$-\frac{Y s_W^2}{24 \sqrt{2}}$ &
$- \frac{Y s_W^2}{32 \sqrt{2}}$ 
\\
$f \bar{f} \rightarrow \gamma \gamma Z$ &
$-\frac{ T_3c_W s_W}{48 \sqrt{2}}$ &  
 $-\frac{ T_3c_W  s_W}{48 \sqrt{2}}$ &
$-\frac{ T_3 c_W s_W}{64 \sqrt{2}}$ &
$-\frac{ T_3c_W s_W}{64 \sqrt{2}}$ &
$-\frac{(1-3s_W^2)( g_L)}{192 \sqrt{2} c_W s_W}$&
${{\frac{(2T_3-Y)s_W c_W}{96 \sqrt{2} }}}$  &  
$-\frac{(\tilde{g})(c_W^2-s_W^2}{96 \sqrt{2} c_W s_W}$&
$-\frac{(1-3s_W^2)( g_L)}{192 \sqrt{2} c_W s_W}$&
$\frac{Y c_W s_W}{24\sqrt{2}}$ &
$\frac{Y c_W s_W}{32\sqrt{2}}$ 
\\
 $f\bar{f} \rightarrow  Z Z Z$ &
$-\frac{ T_3c_W^3}{48 \sqrt{2} s_W}$ &  
 $-\frac{ T_3 c_W^3}{48 \sqrt{2} s_W}$ &
$-\frac{ T_3c_W^3}{64 \sqrt{2} s_W}$ &
$-\frac{ T_3 c_W^3}{64 \sqrt{2} s_W}$ &
${{-\frac{(2T_3-Y)c_W s_W}{128\sqrt{2}}}}$  &
${{-\frac{(2T_3-Y)c_W s_W}{96\sqrt{2}}}}$  &
${{-\frac{(2T_3-Y)c_W s_W}{96\sqrt{2}}}}$  &
${{-\frac{(2T_3-Y)c_W s_W}{128\sqrt{2}}}}$  & 
$\frac{Y s_W^3}{24\sqrt{2} c_W}$ &
$\frac{Y s_W^3}{32\sqrt{2} c_W}$  
\\
$f\bar{f} \rightarrow W^- W^+  \gamma$ &
$0$ &
$-\frac{ T_3}{96 \sqrt{2}}$ &
$ -\frac{ T_3}{192 \sqrt{2}}$ &
$ - \frac{ T_3}{192 \sqrt{2}}$ &
$ -\frac{Y}{384 \sqrt{2}}$ &
$ 0$ &
$-\frac{Y}{192 \sqrt{2}}$ &
$ - \frac{Y}{384 \sqrt{2}}$ &
$ 0$ & $ 0$
\\
 $f\bar{f} \rightarrow W^- W^+ Z$ &
$0$ &
 $ -\frac{ T_3c_W}{96 \sqrt{2} s_W}$ &
$  -\frac{ T_3c_W }{192 \sqrt{2} s_W}$ &
$  -\frac{ T_3c_W }{192 \sqrt{2} s_W}$ &
$ \frac{Ys_W}{384\sqrt{2} c_W}$ &
$ 0$ &
$ \frac{Y s_W}{192\sqrt{2} c_W}$ &
$ \frac{Y s_W}{384\sqrt{2} c_W}$ &
$ 0$ &$0$
\\
 $f_d\,\bar{f}_u \rightarrow W^- W^+ W^-$ & 
 $ 0$ &
 $ -\frac{1}{96 s_W}$ &
$ -\frac{1}{192 s_W}$ &
$ -\frac{1}{192 s_W}$ &
$0$ & $ 0$ & $ 0$ & $ 0$ & $ 0$ & $ 0$
\\
$f_d\,\bar{f}_u \rightarrow \gamma \gamma W^-$ & 
$ 0$ &
$ -\frac{ s_W}{192}$ &
$ -\frac{ s_W}{384}$ &
$ -\frac{ s_W}{384}$ &
$ -\frac{c_W^2}{384 s_W}$ &
$ 0$ &
$ -\frac{c_W^2}{192 s_W}$ &
$ -\frac{c_W^2}{384 s_W}$ &
$ 0$ & $ 0$
\\
 $f_d\,\bar{f}_u \rightarrow  Z Z  W^-$&
 $ 0$ &
 $ -\frac{c_W^2 }{192 s_W}$ &
$ -\frac{c_W^2 }{384 s_W}$ &
$ -\frac{c_W^2 }{384 s_W}$ &
$ -\frac{ s_W}{384}$ &
$ 0$ &
$ -\frac{ s_W}{192}$ &
$ -\frac{ s_W}{384}$ &
$ 0$ & $ 0$
\\
 $f_d\,\bar{f}_u \rightarrow \gamma Z W^- $ & 
 $ 0$ &
 $ -\frac{c_W }{192}$ &
$ -\frac{c_W }{384}$ &
$ -\frac{c_W }{384}$ &
$ \frac{c_W }{384}$ &
$ 0$ &
$ \frac{c_W }{192}$ &
$ \frac{c_W }{384}$ &
$ 0$ & $ 0$
\\
\hline
\end{tabular}
}
\caption{Partial-wave non-orthogonal projections for
  $FF \rightarrow VVV$ generated by ${\cal O}_{T,i}$ with final
  polarizations $++-$ (and permutations) . Here, $f=u,d,\ell$, or
  $\nu_\ell$, while we denote by $f_{u(d)}=u (d)$ or
  $\nu_\ell(\ell)$. The scattering amplitude dependence on the chosen
  fermion is encoded in the couplings
  $g_L=T_{3}- Q s_W^2=T_3c_W^2-\frac{Y}{2}s_W^2$ with the hypercharge
  $Y=2(Q-T_3)$, and the we have defined the combination
  $\tilde{g}\equiv T_3 c_W^2+\frac{Y}{2}s_W^2$. The amplitude for the
  CC processes are the same for quarks and leptons up to the CKM
  factor which we are omitting for simplicity.}
\label{tab:VVVft1}
\end{table}

\begin{table}[H]
\scalebox{0.65}{
\begin{tabular}{|l|c|cccccccccc|}\hline
    & & \multicolumn{10}{c|}
  {$\tilde{T}^{1,a}_{-\hf,\hf;1,1,1}$ in units of 
    $\frac{e}{\pi^2}\frac{1}{576} \sqrt{\frac{70}{31}}      \times
        \frac{f_{T,i}}{\Lambda^4} \, S_{VVV}^2
\times \frac{1}{\sqrt{N_{id}!}}$}\\\hline
        &  $i$   & 0 & 1 & 2 & 3 &4 & 5 &6 &7 &8 &9\\\hline
      & $a$ & \multicolumn{10}{c|}{} \\\hline
$f\bar{f} \rightarrow \gamma \gamma \gamma$     
  & 1 &
  $ -2T_3s_W^2$ &  $-2T_3s_W^2$ & $-\frac{T_3}{2} s_W^2$ & $-\frac{T_3}{2} s_W^2$ &
  $-\frac{\tilde{g}}{2}$ &$-2\tilde{g}$  & $-2\tilde{g}$
  &$-\frac{\tilde{g}}{2}$ &  $-4 Y c_W^2$  & $-Yc_W^2$ \\
  & $ 2 $ &
  $ -2T_3 s_W^2 $ & $ -2T_3 s_W^2 $ & $ -\frac{T_3}{2} s_W^2 $ & $ -\frac{T_3}{2} s_W^2 $ & $ 
  -\frac{\tilde{g}}{2} $ & $-2\tilde{g} $ & $-2\tilde{g}  $ &
  $-\frac{\tilde{g}}{2} $ & $ -4 Yc_W^2  $ & $ -Y c_W^2$ \\
  & $ 3 $ &
  $ 2T_3 s_W^2 $ & $ 2T_3 s_W^2 $ & $ \frac{T_3}{2}  s_W^2 $ & $- \frac{T_3}{2} s_W^2 $ &
  $ \frac{\tilde{g}}{2}  $ & $ 2\tilde{g} $ & $ 2\tilde{g} $ &
  $ \frac{\tilde{g}}{2}  $ & $ 4 Yc_W^2  $ & $ Yc_W^2$ \\
   & $ 4 $&
  $ -2T_3 s_W^2 $ & $ -2T_3 s_W^2 $ & $ -\frac{T_3}{2} s_W^2 $ & $ -\frac{T_3}{2} s_W^2 $ &
  $-\frac{\tilde{g}}{2} $ & $-2\tilde{g} $ & $-2\tilde{g} $ &
  $-\frac{\tilde{g}}{2} $ & $ -4 Yc_W^2  $ & $-Y c_W^2$ \\
   &  5 & 0 & 0 & 0 & 0 & 0 & 0 & 0 & 0 & 0 & 0\\
\hline
$f\bar{f} \rightarrow \gamma Z Z$
& 1 &
$ -2T_3 c_W^2  $ & $ -2T_3c_W^2  $ & $- \frac{T_3}{2} c_W^2$ & $ -\frac{T_3}{2} c_W^2$ &
$ \frac{Q+5\tilde{g}}{10} $ & $ \frac{-6Q+10\tilde{g}}{5} $
& $ -\frac{2Q-10\tilde{g}}{5} $ &
$ \frac{-3Q+5\tilde{g}}{10} $ &
$ -4Y  s_W^2 $ & $-Y  s_W^2$\\
& 2 &
$ -2T_3 c_W^2  $ & $ -2T_3 c_W^2  $ & $-\frac{T_3}{2} c_W^2$ & $ -\frac{T_3}{2} c_W^2$ &
$ \frac{Q+5\tilde{g}}{10} $ & $ \frac{-6Q+10\tilde{g}}{5} $
& $ -\frac{2Q-10\tilde{g}}{5} $ &
$ \frac{-3Q+5\tilde{g}}{10} $ &
$  -4 Y s_W^2 $ & $ -Y s_W^2$\\
& 3 &
$ -2T_3 c_W^2  $ & $ 2T_3c_W^2$ & $ \frac{T_3}{2} c_W^2 $ & $ -\frac{T_3}{2} c_W^2  $ &
$ \frac{Q-\tilde{g}}{2}  $ & $-(2\tilde{g})$ &
$ {{-\frac{(2T_3-Y)(c_W^2-s_W^2)}{2}}}$ & $-\frac{\tilde{g}}{2} $ & $ 4 Y s_W^2 $ & $ Ys_W^2$\\
& 4 &
$ -2T_3c_W^2  $ & $ -2T_3 c_W^2  $ & $ -\frac{T_3}{2} c_W^2$ & $ -\frac{T_3}{2} c_W^2$ &
$ -\frac{Q-5\tilde{g}}{10} $ &
$ -\frac{4Q-10\tilde{g}}{5}  $ &
$-\frac{6Q-20\tilde{g}}{10} $ &
$ -\frac{2Q-5\tilde{g}}{10}  $ & $ -4 Y s_W^2 $ & $ -Y s_W^2$\\
& 5 &
$ -2T_3c_W^2  $ & $ 0 $ & $ 0 $ & $ 0 $ & $ \frac{Q}{5}\sqrt{\frac{31}{10}}  $ & $-\frac{Q}{5} \sqrt{\frac{31}{10}} $
& $ \frac{Q}{5}\sqrt{\frac{31}{10}}$ & $-\frac{Q}{10}\sqrt{\frac{31}{10}} $ & $ 0 $  & $ 0$\\
\hline
$f\bar{f} \rightarrow \gamma \gamma Z$
&1 &
$ -2T_3 c_W  s_W $ & $-2T_3 c_W  s_W $ & $ -\frac{T_3 c_W  s_W}{2} $ & $ -\frac{T_3 c_W  s_W}{2} $ &
$- \frac{2 g_L-5s_W^2c_W^2(2T_3 -Y)}{20 c_W s_W} $
& $- \frac{4 g_L-5s_W^2c_W^2(2T_3 -Y) }{5 s_W c_W} $
& $- \frac{3 g_L-5s_W^2c_W^2(2T_3 -Y) }{5 s_W c_W} $ &
$  \frac{4 g_L-5 s_W^2c_W^2(2T_3 -Y)}{20 c_W s_W} $
& $ 4Y c_W  s_W $ & $ Y c_W  s_W$ \\ 
& 2 &
$-2T_3 c_W  s_W $ & $ -2T_3c_W  s_W $ & $ -\frac{T_3c_W  s_W}{2} $ & -$\frac{T_3 c_W  s_W}{2} $ &
$ \frac{2T_3c_W^4+Y s_W^4}{4 s_W c_W} $ &
$ {{(2T_3-Y) s_W c_W}} $& $-\frac{(c_W^2-s_W^2)(2\tilde{g}) }{2 s_W c_W} $ &
$ {{\frac{(2T_3-Y)s_W c_W}{4}}} $& $ 4Y c_W  s_W $ & $ Y c_W  s_W $ \\    
& 3  &
$ 2T_3 c_W  s_W $ & $ 2T_3 c_W  s_W  $ & $ \frac{T_3c_W  s_W}{2} $ & $ \frac{T_3c_W  s_W}{2} $ &
$ \frac{2 g_L-5s_W^2c_W^2(2T_3 -Y)}{20 c_W s_W} $
& $ \frac{4 g_L-5s_W^2c_W^2(2T_3 -Y)}{5 s_W c_W} $
& $ \frac{3 g_L-5s_W^2c_W^2(2T_3 -Y) }{5 s_W c_W} $ &
$  -\frac{4 g_L-5s_W^2c_W^2(2T_3 -Y)}{20 c_W s_W} $
& $ -4Y c_W  s_W $ & $ -Yc_W  s_W$ \\ 
&4   &
$ -2T_3c_W  s_W $ & $ -2T_3 c_W  s_W $ & $ -\frac{T_3 c_W  s_W}{2} $ & $  -\frac{T_3c_W  s_W}{2} $ &
$ \frac{2T_3c_W^4+Y s_W^4}{4 s_W c_W} $ &
$ {{(2T_3-Y) s_W c_W}} $ & $-\frac{(c_W^2-s_W^2)(\tilde{g}) }{ s_W c_W} $ &
$ {{\frac{(2T_3-Y)s_W c_W}{4}}} $ & $ 4 Y c_W  s_W $ & $ Y c_W  s_W$ \\ 
&5  & $ 0 $ & $ 0 $ & $ 0 $ & $ 0 $ & $ -\sqrt{\frac{31}{10}} \frac{2 g_L}{10 c_W s_W} $ &
$ \sqrt{\frac{31}{10}}   \frac{2 g_L}{5 c_W s_W} $ & $ -\sqrt{\frac{31}{10}}  \frac{2 g_L}{10 c_W s_W} $ &
$ \sqrt{\frac{31}{10}}  \frac{2 g_L}{20 c_W s_W} $ & $ 0 $ & $ 0$ \\
\hline
 $f\bar{f} \rightarrow  Z Z Z$ 
&1  &
$ -\frac{2T_3c_W^3}{s_W} $ & $ -\frac{2T_3c_W^3}{s_W} $ & $ -\frac{T_3c_W^3}{2 s_W} $ & $-\frac{T_3c_W^3}{2 s_W}$ &
$ {{-\frac{(2T_3-Y)s_W c_W}{4}}} $ & $ {{-(2T_3-Y)s_W c_W}} $ & $ {{-(2T_3-Y)s_W c_W}} $ 
& $ {{-\frac{(2T_3-Y)s_W c_W}{4}}} $
& $ 4 \frac{Y s_W^3}{c_W} $& $  -\frac{Y s_W^3}{c_W}$ \\ 
&2  &
$ -\frac{2T_3c_W^3}{s_W} $ & $-\frac{2T_3c_W^3}{s_W} $ & $-\frac{T_3c_W^3}{2 s_W} $ & $-\frac{T_3c_W^3}{2 s_W} $
& $ {{-\frac{(2T_3-Y)s_W c_W}{4}}} $ & $ {{-(2T_3-Y)s_W c_W}} $ & $ {{-(2T_3-Y)s_W c_W}} $ 
& $ {{-\frac{(2T_3-Y)s_W c_W}{4}}} $
& $ 4 \frac{Y s_W^3}{c_W} $ & $ - \frac{Ys_W^3}{c_W}$ \\ 
&3  &
$ \frac{2T_3c_W^3}{s_W}  $ & $ \frac{2T_3c_W^3}{s_W}  $ & $\frac{T_3c_W^3}{2 s_W} $ & $\frac{T_3c_W^3}{2 s_W} $ &
$ {{\frac{(2T_3-Y)s_W c_W}{4}}} $ & $ {{(2T_3-Y)s_W c_W}} $ & $ {{(2T_3-Y)s_W c_W}} $ 
& $ {{\frac{(2T_3-Y)s_W c_W}{4}}} $
& $ -4 \frac{Ys_W^3}{c_W} $ & $ -\frac{Ys_W^3}{c_W}$ \\ 
&4  &
$-\frac{2T_3c_W^3}{s_W} $ & $-\frac{2T_3c_W^3}{s_W} $ & $-\frac{T_3c_W^3}{2 s_W} $ & $ -\frac{T_3c_W^3}{2 s_W} $ &
$ {{-\frac{(2T_3-Y)s_W c_W}{4}}} $ & $ {{-(2T_3-Y)s_W c_W}} $ & $ {{-(2T_3-Y)s_W c_W}} $ 
& $ {{-\frac{(2T_3-Y)s_W c_W}{4}}} $
& $ 4 \frac{Ys_W^3}{c_W} $ & $  -\frac{Ys_W^3}{c_W}$ \\ 
&5  & $ 0 $ & $ 0 $ & $ 0 $ & $ 0 $ & $ 0 $ & $ 0 $ & $ 0 $ & $ 0 $ & $ 0 $ & $ 0$ \\
\hline
 $f\bar{f} \rightarrow  W^+ W^- \gamma$ 
&1  &
$ -\frac{2T_3}{5} $ & $ -\frac{3T_3}{5} $ & $-\frac{T_3}{5} $ & $ -\frac{T_3}{10} $ &
$ -\frac{Y}{20} $ & $ -\frac{2Y}{5} $ & $ -\frac{3Y}{10} $ & $ -\frac{Y}{10} $ & $ 0 $ & $ 0$ \\ 
&2 &
$ 0 $ & $-T_3$ & $ 0 $ & $-\frac{T_3}{2} $ &
$-\frac{Y}{4} $ & $ 0 $ & $ -\frac{Y}{2} $ & $ 0 $ & $ 0 $ & $ 0$ \\  
&3 &
$ \frac{4T_3}{5} $ & $ \frac{3T_3}{5} $ & $ \frac{T_3}{5}  $ & $ \frac{T_3}{10}  $ &
$ \frac{Y}{20}  $ & $ \frac{2Y}{5} $ & $\frac{3Y}{10} $ & $ \frac{Y}{10}  $ & $ 0 $ & $ 0$ \\ 
&4 &
$ 0 $ & $ -T_3 $ & $ 0 $ & $-\frac{T_3}{2} $ &
$-\frac{Y}{4} $ & $ 0 $ & $ -\frac{Y}{2} $ & $ 0 $ & $0$ & $0$ \\
&5 &
$ \frac{2T_3}{5} \sqrt{\frac{31}{10}} $ & $-\frac{T_3}{5} \sqrt{\frac{31}{10}} $ & $ \frac{T_3}{10} \sqrt{\frac{31}{10}} $
& $-\frac{T_3}{5} \sqrt{\frac{31}{10}} $ &
$-\frac{Y}{10} \sqrt{\frac{31}{10}} $ & $\frac{Y}{5} \sqrt{\frac{31}{10}} $ & $-\frac{Y}{10} \sqrt{\frac{31}{10}} $ &
$ \frac{Y}{20} \sqrt{\frac{31}{10}} $ & $ 0 $ & $0$ \\
\hline
 $f\bar{f} \rightarrow  W^+ W^- Z$ 
&1 &
$ -\frac{4T_3 c_W }{5 s_W}$ & $- \frac{3T_3 c_W }{5 s_W}$ & $ -\frac{T_3 c_W }{5 s_W}$ & $-\frac{T_3c_W }{10 s_W}$ &
$ \frac{Y s_W}{20 c_W}$ & $ \frac{2Y  s_W}{5 c_W}$ & $ \frac{3 Y s_W}{10 c_W}$ & $ \frac{Y s_W}{10 c_W}$ & $ 0$ & $ 0$ \\
&2 &
$ 0$ & $ -\frac{T_3c_W }{ s_W}$ & $ 0$ & $ -\frac{T_3c_W }{2 s_W}$ &
$ \frac{Y s_W}{4 c_W}$ & $ 0$ & $ \frac{Y s_W}{2 c_W}$ & $ 0$ & $ 0$ & $ 0$ \\
&3 &
$ \frac{4T_3 c_W }{5 s_W}$ & $ \frac{3T_3 c_W }{5 s_W}$ & $\frac{T_3c_W }{5 s_W}$ & $ \frac{T_3c_W }{10 s_W}$ &
$- \frac{Y s_W}{20 c_W}$ & -$Y \frac{2  s_W}{5 c_W}$ & $-\frac{3Y  s_W}{10 c_W}$ & $ -\frac{Y s_W}{10 c_W}$ & $ 0$ & $ 0$ \\
&4  &
$ 0$ & $ -\frac{T_3c_W }{s_W}$ & $ 0$ & $ -\frac{T_3c_W }{2 s_W}$ &
$ \frac{Y s_W}{4c_W}$ & $ 0$ & $ \frac{Y s_W}{2c_W}$ & $ 0$ & $ 0$ & $ 0$ \\
&5 &
$\sqrt{\frac{31}{10}} \frac{2T_3 c_W }{5 s_W}$ & $-\sqrt{\frac{31}{10}} \frac{T_3 c_W }{5 s_W}$ &
$ \sqrt{\frac{31}{10}} \frac{T_3c_W }{10 s_W}$ & $-\sqrt{\frac{31}{10}} \frac{T_3 c_W }{5 s_W}$ &
$ \sqrt{\frac{31}{10}}\frac{Y  s_W}{10 c_W}$ & $-\sqrt{\frac{31}{10}} \frac{Y s_W}{5 c_W}$ &
$ \sqrt{\frac{31}{10}} \frac{Y s_W}{10 c_W}$ &$ -\sqrt{\frac{31}{10}} \frac{Y s_W}{20 c_W}$ & $ 0$ & $ 0$ \\
   \hline
$f_d\,\bar{f}_u \rightarrow W^- W^+ W^-$
  & 1  & $ -\frac{2 \sqrt{2} }{5 s_W}$ & $ -\frac{4 \sqrt{2}}{5 s_W}$ & $ -\frac{1}{5\sqrt{2} s_W}$ & $ -\frac{3}{5 \sqrt{2} s_W}$ & $ 0$ & $ 0$ & $ 0$ & $ 0$ & $ 0$ & $ 0$ \\ 
  & 2  & $ -\frac{2 \sqrt{2} }{5 s_W}$ & $ -\frac{4 \sqrt{2}}{5 s_W}$ & $ -\frac{1}{5\sqrt{2} s_W}$ & $ -\frac{3}{5 \sqrt{2} s_W}$ & $ 0$ & $ 0$ & $ 0$ & $ 0$ & $ 0$ & $ 0$ \\ 
  & 3  & $ \frac{\sqrt{2}}{s_W}$ & ${1}{\sqrt{2} s_W}$ & $\frac{1}{2 \sqrt{2} s_W}$ & $ 0$ & $ 0$ & $ 0$ & $ 0$ & $ 0$ & $ 0$ & $ 0$ \\
  & 4  & $ -\frac{3 \sqrt{2} }{5 s_W}$ & $ -\frac{7 }{5 \sqrt{2} s_W}$ & $ -\frac{3 }{10 \sqrt{2} s_W}$ & $ -\frac{\sqrt{2} }{5s_W}$ & $ 0$ & $ 0$ & $ 0$ & $ 0$ & $ 0$ & $ 0$ \\ 
  & 5  & $\sqrt{\frac{31}{5}}\frac{1}{5 s_W}$ & $ -\sqrt{\frac{31}{5}}\frac{1}{10 s_W}$ & $ \sqrt{\frac{31}{5}}\frac{ 1}{20 s_W}$ & $ -\sqrt{\frac{31}{5}}\frac{1}{10 s_W}$ & $ 0$ & $ 0$ & $ 0$ & $ 0$ & $ 0$ & $ 0$ \\   
\hline
 $f_d\,\bar{f}_u \rightarrow \gamma\gamma W^-$ 
& 1 & $ -\frac{2 \sqrt{2}  s_W}{5}$ & $ -\frac{3  s_W}{5 \sqrt{2}}$ & $ -\frac{s_W}{5\sqrt{2}}$ & $-\frac{ s_W}{10\sqrt{2}}$ & $ -\frac{c_W^2}{10\sqrt{2} s_W}$ &
$ -\frac{2 \sqrt{2} c_W^2 }{5 s_W}$ & $ -\frac{3 c_W^2 }{5 \sqrt{2} s_W}$ & $ -\frac{c_W^2}{5\sqrt{2} s_W}$ & $ 0$ & $ 0$\\
& 2 & $ 0$ & $-\frac{s_W}{\sqrt{2}}$ & $ 0$ & $-\frac{s_W}{2\sqrt{2}}$ & $ -\frac{c_W^2}{2\sqrt{2} s_W}$ & $ 0$ & $-\frac{c_W^2 }{\sqrt{2} s_W}$ & $ 0$ & $ 0$ & $ 0$\\
& 3 & $ \frac{2 \sqrt{2}  s_W}{5}$ & $\frac{3  s_W}{5 \sqrt{2}}$ & $ \frac{ s_W}{5 \sqrt{2}}$ & $ \frac{ s_W}{10 \sqrt{2}}$ &
$\frac{c_W^2 }{10 \sqrt{2} s_W}$ & $ \frac{2 \sqrt{2} c_W^2}{5 s_W}$ & $\frac{3 c_W^2 }{5 \sqrt{2} s_W}$ & $ \frac{c_W^2}{5 \sqrt{2} s_W}$ & $ 0$ & $ 0$\\ 
  & 4 & $ 0$ & $ -\frac{ s_W}{\sqrt{2}}$ & $ 0$ & $ -\frac{ s_W}{2\sqrt{2}}$ & $-\frac{c_W^2 }{2\sqrt{2} s_W}$ & $ 0$ & $ -\frac{c_W^2 }{\sqrt{2} s_W}$ & $ 0$ & $ 0$ & $ 0$\\ 
   & 5 & $ \sqrt{\frac{31}{5}}\frac{ s_W}{5}$ & $ -\sqrt{\frac{31}{5}}\frac{ s_W}{10}$ &
   $ \sqrt{\frac{31}{5}}\frac{s_W}{20}$ & $ -\sqrt{\frac{31}{5}}\frac{ s_W}{10}$ & $-\sqrt{\frac{31}{5}}\frac{10 c_W^2 }{s_W}$ & $ \sqrt{\frac{31}{5}}\frac{ c_W^2 }{5 s_W}$ &
   $ -\sqrt{\frac{31}{5}}\frac{ 10c_W^2 }{s_W}$ & $ \sqrt{\frac{31}{5}}\frac{ c_W^2 }{20 s_W}$ & $ 0$ & $ 0$\\ 
\hline
 $f_d\,\bar{f}_u \rightarrow  Z Z W^-$ 
& 1  & $ -\frac{2 \sqrt{2} c_W^2}{5 s_W}$ & $ -\frac{3 c_W^2 }{5 \sqrt{2} s_W}$ & $ -\frac{c_W^2}{5\sqrt{2} s_W}$ & $ -\frac{c_W^2 }{10\sqrt{2} s_W}$ &
$ -\frac{s_W}{10\sqrt{2}}$ & $ -\frac{2 \sqrt{2} s_W}{5}$ & $ -\frac{3 s_W}{5 \sqrt{2}}$ & $-\frac{s_W}{5\sqrt{2}}$ & $ 0$ & $ 0$\\
& 2  & $ 0$ & $ -\frac{c_W^2}{\sqrt{2} s_W}$ & $ 0$ & $-\frac{c_W^2}{2\sqrt{2} s_W}$ & $ -\frac{s_W}{2\sqrt{2}}$ & $ 0$ & $-\frac{s_W}{\sqrt{2}}$ & $ 0$ & $ 0$ & $ 0$\\
& 3  & $ \frac{2 \sqrt{2} c_W^2 }{5 s_W}$ & $\frac{3 c_W^2 }{5 \sqrt{2} s_W}$ & $ \frac{c_W^2 }{5 \sqrt{2} s_W}$ & $\frac{c_W^2 }{10 \sqrt{2} s_W}$ &
$ \frac{s_W}{10 \sqrt{2}}$ & $ \frac{2 \sqrt{2}  s_W}{5}$ & $\frac{3 s_W}{5 \sqrt{2}}$ & $ \frac{ s_W}{5 \sqrt{2}}$ & $ 0$ & $ 0$\\ 
  & 4  & $ 0$ & $ -\frac{c_W^2 }{\sqrt{2} s_W}$ & $ 0$ & $ -\frac{c_W^2}{2\sqrt{2} s_W}$ & $-\frac{s_W}{2\sqrt{2}}$ & $ 0$ & $ -\frac{ s_W}{\sqrt{2}}$ & $ 0$ & $ 0$ & $ 0$\\ 
& 5  & $ \sqrt{\frac{31}{5}}\frac{ c_W^2 }{5 s_W}$ & $ -\sqrt{\frac{31}{5}}\frac{ c_W^2 }{10 s_W}$ & $\sqrt{\frac{31}{5}}\frac{ c_W^2 }{20 s_W}$ &
$ -\sqrt{\frac{31}{5}}\frac{ c_W^2}{10 s_W}$ & $ -\sqrt{\frac{31}{5}}\frac{  s_W}{10}$ & $ \sqrt{\frac{31}{5}}\frac{  s_W}{5}$ & $ -\sqrt{\frac{31}{5}}\frac{  s_W}{10}$ &
$ \sqrt{\frac{31}{5}}\frac{  s_W}{20}$ & $ 0$ & $ 0$\\
\hline
 $f_d\,\bar{f}_u \rightarrow \gamma Z W^-$  
 & 1 & $ -\frac{2 \sqrt{2} c_W }{5}$ & $ -\frac{3 c_W }{5 \sqrt{2}}$ & $ - \frac{c_W }{5\sqrt{2}}$ & $ 
   -\frac{c_W}{10\sqrt{2}}$ & $ \frac{c_W}{10 \sqrt{2}}$ & $ \frac{2 \sqrt{2} c_W }{5}$ & $\frac{3 c_W}{5 \sqrt{2}}$ & $ \frac{c_W}{5 \sqrt{2}}$ & $ 0$ & $ 0$\\ 
  & 2 & $ 0$ & $ -\frac{c_W}{\sqrt{2}}$ & $ 0$ & $ -\frac{c_W}{2\sqrt{2}}$ & $ \frac{c_W}{2 \sqrt{2}}$ & $    0$ & $ \frac{c_W }{\sqrt{2}}$ & $ 0$ & $ 0$ & $ 0$\\
  & 3 & $ \frac{2 \sqrt{2} c_W }{5}$ & $ \frac{3 c_W}{5 \sqrt{2}}$ & $ \frac{c_W}{5 \sqrt{2}}$ & $ \frac{c_W}{10 \sqrt{2}}$ & $ 
   -\frac{c_W}{10\sqrt{2}}$ & $ -\frac{2 \sqrt{2} c_W }{5}$ & $ -\frac{3 c_W }{5 \sqrt{2}}$ & $ 
   -\frac{c_W}{5\sqrt{2}}$ & $ 0$ & $ 0$\\
   & 4 & $ 0$ & $ -\frac{c_W }{\sqrt{2}}$ & $ 0$ & $ 
   -\frac{c_W}{2\sqrt{2}}$ & $ \frac{c_W}{2 \sqrt{2}}$ & $ 0$ & $ \frac{c_W}{\sqrt{2}}$ & $ 0$ & $ 0$ & $    0$\\
   & 5 & $ \sqrt{\frac{31}{5}}\frac{ c_W}{5}$ & $ -\sqrt{\frac{31}{5}}\frac{ c_W}{10}$ & $ 
   \sqrt{\frac{31}{5}}\frac{ c_W}{20}$ & $ -\sqrt{\frac{31}{5}}\frac{ c_W}{10}$ & $ \sqrt{\frac{31}{5}}\frac{ c_W }{10}$ & $ 
   -\sqrt{\frac{31}{5}}\frac{ c_W}{5}$ & $ \sqrt{\frac{31}{5}}\frac{ c_W }{10}$ & $ 
   -\sqrt{\frac{31}{5}}\frac{ c_W }{20}$ & $ 0$ & $ 0$\\
\hline
\end{tabular}}
\caption{Partial-wave non-orthogonal projections for
  $FF \rightarrow VVV$ generated by ${\cal O}_{T,i}$ with final
  polarizations $+++$. Conventions are the same as in
  Table~\ref{tab:VVVft1}.  }
\label{tab:VVVft2}
\end{table}

\newpage
\bibliography{quarticreferences}

\end{document}